\newif\ifpretty
\prettytrue

\ifpretty
\documentclass[pra,twocolumn,preprintnumbers,amsmath,amssymb]{revtex4}
\else
\documentclass[prb,preprint,showpacs,preprintnumbers,amsmath,amssymb]{revtex4}
\fi

\newif\ifpdf
        \ifx\pdfoutput\undefined
        \pdffalse 
        \else
        \pdfoutput=1 
        \pdftrue
        \fi
\ifpdf
   \usepackage[pdftex]{graphicx}
   \DeclareGraphicsExtensions{.pdf, .png}
   \graphicspath{{./FIG/}}
   \usepackage{thumbpdf}
\else
   \usepackage{graphicx}
   \DeclareGraphicsExtensions{.eps}
   \graphicspath{{./FIG/}}
\fi

\begin{document}

\title{
Heat Capacity Effects Associated with the
Hydrophobic Hydration and Interaction of Simple Solutes:
A Detailed Structural and Energetical Analysis Based
on MD Simulations}

\author{Dietmar Paschek}
\email{dietmar.paschek@udo.edu}
\affiliation{Department of Physical Chemistry, Otto-Hahn Str. 6,
  University of Dortmund, D-44221 Dortmund, Germany}


\date{\today}

\begin{abstract}
We examine the SPCE and TIP5P water models
using a temperature series of 
MD simulations in order to study heat capacity effects
associated with the hydrophobic hydration and interaction
of Xenon particles. The temperature interval
between $275\,\mbox{K}$ and $375\,\mbox{K}$ along the $0.1\;\mbox{MPa}$ isobar
is studied.
For all investigated models and state points
we calculate the excess chemical potential for Xenon
employing the Widom particle insertion technique.
The solvation enthalpy and excess heat capacity is obtained from the temperature
dependence of the chemical
potentials and, alternatively, directly by Ewald summation, as well as
a reaction field based method. All three different 
approaches provide consistent results.
In addition, the employed reaction field method allows a separation of
the individual components to the solvation enthalpy into solute/solvent
and solvent/solvent parts. We find that the solvent/solvent contribution
to the excess heat capacity is the dominating contribution, being about
one order of magnitude larger than the solute/solvent part. The latter contribution
is found to be due to the enlarged heat capacity of the water molecules in the
hydration shell.
A detailed spacial analysis of the heat capacity of the water molecules around
a pair of Xenon particles at different separations reveals that the
particularly enhanced heat capacity of the
water molecules in the bisector plane between two
Xenon atoms is responsible
for the maximum of the heat capacity observed at the desolvation barrier,
recently reported by Shimizu and Chan
({\em J. Am. Chem. Soc.},{\bf 123}, 2083--2084 (2001)).
The about 60\% enlarged heat capacity of water in the concave part of 
the joint Xenon-Xenon hydration shell 
is found to be the result of a counterplay of strengthened 
hydrogen bonds and an enhanced
breaking of hydrogen bonds with increasing temperature.
Differences between the two models concerning the heat capacity in
the Xenon-Xenon contact state are attributed to the
different water model bulk heat capacities, and
to the different spacial extension
of the structure effect introduced by the hydrophobic particles.
Similarities between the different states of water in the joint 
Xenon-Xenon hydration shell and the properties of stretched water 
are discussed.
\end{abstract}


\maketitle

\section{INTRODUCTION}

\ifpretty\enlargethispage{1em}\fi
Nonpolar simple solutes, such as noble gases or methane, don't like to be
dissolved in water: They are hydrophobic. Hence the corresponding 
solvation free energy is found to be 
large and positive \cite{Tanford,Ben-Naim:Hydrophobic,Southall:2002}.
However, the temperature dependence of the hydration free energy
of hydrophobic solutes reveals a {\em negative} solvation enthalpy,
which is counterbalanced by an
also {\em negative} solvation entropy
\cite{Tanford,Ben-Naim:Hydrophobic,Southall:2002}. 
For the case of the simple solutes the latter observation has
to be attributed to structural organization
 processes in the
solvent phase, due to the lack of the solutes internal degrees of freedom.
In addition, the solvation excess heat capacity is found to be {\em positive}, 
typically being in the range between $150\,\mbox{J}\,\mbox{K}^{-1}\,\mbox{mol}^{-1}$
and $300\,\mbox{J}\,\mbox{K}^{-1}\,\mbox{mol}^{-1}$
around $300\,\mbox{K}$ \cite{Southall:2002}, strongly
suggesting an enhanced heat capacity of the hydration shell water.
This essential feature has been used as
a key ingredient to model hydrophobic hydration effects 
on the basis of a solvent structural (basically ``hydrogen bond'') reorganization 
\cite{Muller:90,LeeB:95,Moelbert:2003:2,Moelbert:2003:1}.

As a consequence of overlapping hydration shells and the release of
{\em shell} water molecules into the {\em bulk} phase,
the association of hydrophobic particles is found to be strengthened
with increasing temperature
in order to minimize the solvation entropy penalty \cite{Smith.D:92,Smith.D:93}.
This sort of entropy-driven association process is usually referred to as
{\em hydrophobic interaction} and has been qualitatively confirmed by numerous
simulation studies
\cite{Geiger:79,Zichi:85,Smith.D:92,Smith.D:93,Pearlman:93,%
Belle:93,Dang:94,Forsman:94,Luedemann:96,Skipper:96,Young:97,%
Hummer:96:1,Luedemann:97,Rick:97,Hummer:2001,%
Shimizu:2000,Rick:2000,Shimizu:2001,Shimizu:2002,%
Ghosh:2001,Ghosh:2002,Ghosh:2003,Rick:2003}.
In line with the negative net entropy of association, also a 
negative heat capacity contribution is expected and has been
incorporated into models describing the hydrophobic association
on the basis of the solvent exposed surface area \cite{Lee:71,Makhatadze:95}.
For a full depth review on the present status of conceptual understanding of
hydrophobic effects we would like to refer to the recent
review articles
of Southall et al. \cite{Southall:2002}, Pratt \cite{PrattRev:2002}, 
and Widom \cite{Widom:2003}. In addition, Smith and Haymet 
\cite{Smith:2003} provide a tutorial
overview over related simulation techniques.

However, recent publications by Shimizu and Chan
\cite{Shimizu:2000,Shimizu:2001,Shimizu:2002}, Rick \cite{Rick:2003} 
and Paschek \cite{Paschek:2004} indicate that the heat capacity change,
related to the association of two hydrophobic particles,
exhibits a maximum located at distance of the
desolvation barrier around $5.5-6.5\,\mbox{\AA}$.
The presence and observed strength of the maximum has not been considered 
in models of the hydrophobic interaction 
based on the solvent
accessible surface \cite{Shimizu:2001,Shimizu:2002}
 and has  also consequences for the
heat net capacity change at the contact state.
Nevertheless, some uncertainty exists about the strength
of this effect for hydrophobic particles in the contact state.
Shimizu et al. find an almost vanishing \cite{Shimizu:2001}
contribution to the heat capacity (nearly identical heat capacities
for the contact state and the fully separated state), whereas Rick
\cite{Rick:2003} observes a negative net heat capacity for the
associated state, being qualitatively in accord with model predictions based
on the change of the solvent accessible surface. Both results were
obtained for a pair of methane particles dissolved in TIP4P model water.
Paschek \cite{Paschek:2004} finds evidence for both scenarios, 
depending on the water model which is taken into account:
For the case of the association of Xenon particles
the SPCE model shows the behavior reported by Shimizu et al., whereas
the TIP5P model corresponds to Rick's scenario.

In this paper we will provide a quantitative energetical and structural 
study on the origin
of the observed heat capacity effects. We would
like to show that (at least for the studied model systems) the
solvation heat capacity is dominated by the solvent restructuring
effects and is related to the enhanced breaking
of hydrogen bonds in different parts of the
the hydration shell with increasing temperature.
In addition, we try to elaborate the reason for the observed
differences between the SPCE and TIP5P models.
Finally, we would like show that observed heat capacity effects can
be understood from the properties of {\em stretched} water.


\section{METHODS}

\subsection{MD Simulation details}

\label{sec:MD}

We employ molecular dynamics (MD) simulations in the NPT ensemble using
the Nos\'e-Hoover thermostat 
\cite{Nose:84,Hoover:85}
and the Rahman-Parrinello barostat 
\cite{Parrinello:81,Nose:83} with
coupling times $\tau_T\!=\!1.5\,\mbox{ps}$ and 
$\tau_p\!=\!2.5\,\mbox{ps}$
(assuming the isothermal compressibility to be 
$\chi_T\!=\!4.5\;10^{-5}\,\mbox{bar}^{-1}$), respectively.
The electrostatic interactions are treated
in the ``full potential'' approach
by the smooth particle mesh Ewald summation 
\cite{Essmann:95} with a real space
cutoff of $0.9\,\mbox{nm}$ and a mesh spacing of approximately
$0.12\,\mbox{nm}$ and 4th order
interpolation. The Ewald convergence factor $\alpha$ was set to
$3.38\,\mbox{nm}^{-1}$ (corresponding to a relative accuracy of
the Ewald sum of $10^{-5}$).
A $2.0\,\mbox{fs}$ 
timestep was used for all simulations and the constraints were solved
using the SETTLE procedure \cite{Miyamoto:92}.
All simulations reported here were carried out using 
the GROMACS 3.1  program \cite{gmxpaper,gmx31}.
Statistical errors in the analysis
were computed using the method of Flyvbjerg and Petersen \cite{Flyvbjerg:89}.
For all reported systems and different
statepoints initial equilibration runs of $1\,\mbox{ns}$ length 
were performed using the Berendsen 
weak coupling scheme for pressure and temperature control
($\tau_T\!=\!\tau_p\!=\!0.5\,\mbox{ps}$) \cite{Berendsen:84}. 

In order to obtain the
solvation enthalpies and excess heat capacities of Xenon,
we follow two distinct approaches:
1) Indirectly from the temperature dependence the excess chemical potential.
2) Directly from the energies obtained from the simulation runs.
Since this procedure
is just intended to show the applicability of our approach, it
is applied only to the SPCE model. For this purpose
we performed a series of simulations containing
500 water molecules employing the SPCE model \cite{Berendsen:87} as well
as a series containing 500 SPCE molecules plus an additional Xenon
(Xenon-Xenon parameters: $\sigma\!=\!3.975\,\mbox{\AA}$, 
$\epsilon\,k_B^{-1}\!=\!214.7\,\mbox{K}$ \cite{Guillot:93})
particle.
The systems were simulated at five different temperatures
$275\,\mbox{K}$, $300\,\mbox{K}$, $325\,\mbox{K}$, $350\,\mbox{K}$
and $375\,\mbox{K}$ at a pressure of $0.1\,\mbox{MPa}$. 
Each of these simulations extended to $20\,\mbox{ns}$  and $2\times 10^4$ configurations
were stored for further analysis. 

To determine the heat capacity change for the association
of Xenon particles
we use MD simulations containing 500 water SPCE and TIP5P \cite{Mahoney:2000} molecules and
8 Xenon particles employing the simulation conditions outlined
above. Again, each of the model systems is
studied at $275\,\mbox{K}$, $300\,\mbox{K}$, $325\,\mbox{K}$, $350\,\mbox{K}$
and $375\,\mbox{K}$ at a pressure of $0.1\,\mbox{MPa}$.
Here, runs over $60\,\mbox{ns}$ were conducted, while storing $7.5\times 10^4$ 
configurations for further analysis.
The water/Xenon parameters were obtained applying the standard
Lorentz-Berthelot mixing rules according to
$\sigma_{ij}\!=\!\left(\sigma_{ii}+\sigma_{jj}\right)/2$ and
$\epsilon_{ij}\!=\!\sqrt{\epsilon_{ii}\epsilon_{jj}}$.

\subsection{Hydrophobic Hydration and Interaction}
\label{sec:hydrophobic}

We calculate the excess chemical potential of Xenon in
water {\em a posteriori} from simulation trajectories 
obtained at constant pressure/temperature (NPT-Ensemble) 
conditions. For this purpose we employ
the Widom particle
insertion method \cite{Widom:63,FrenkelSmit} according to
\begin{eqnarray}
  \label{eq:muexdef}
  \mu_{ex}
  & = &
  - \beta^{-1} 
  \ln \frac{\left< V \int d\vec{s}_{N+1} \exp(-\beta\,\Delta
  U\right>}{\left<V\right>} 
\end{eqnarray}
where 
$\Delta U\!=\! U(\vec{s}^{N+1};L)-U(\vec{s}^{N};L)$ 
is the
potential energy of a randomly inserted solute $(N+1)$- particle into
a configuration containing $N$ solvent
molecules. The $\vec{s}_{i}=L^{-1}\vec{r}_{i}$ (with $L\!=\!V^{1/3}$ being the
length of a hypothetical cubic box) are the 
scaled coordinates of the particle positions and $\int\vec{s}_{N+1}$ denotes
an integration over the whole space. 
The brackets $\left<\ldots\right>$ indicate isothermal-isobaric
averaging over the configuration space of the $N$-particle system (the solvent).
The entropic and enthalpic contributions to the excess chemical potential can
be obtained straightforwardly as temperature derivative according to
\begin{equation}
s_{ex}  =  - \left(\frac{\partial \mu_{ex}}{\partial T}\right)_P
\hspace*{2em}\mbox{and}\hspace*{2em}
h_{ex} =   \mu_{ex}+T\,s_{ex}
\end{equation}
and the isobaric heat capacity contribution according to
\begin{eqnarray}
c_{P,ex} & = &  -T \left( \frac{\partial^2\mu_{ex}}{\partial T^2}\right)_P
= \left( \frac{\partial h_{ex}}{\partial T}\right)_P \;.
\end{eqnarray}

In order to perform the calculation most efficiently we have made use of
the excluded volume map (EVM) technique \cite{Deitrick:89,Deitrick:92} by
mapping the occupied volume onto a grid of approximately $0.2\,\mbox{\AA}$ mesh-width. Distances smaller
than $0.7\times\sigma_{ij}$ with respect to any solute molecule (oxygen site)
were neglected and the term $exp(-\beta\,\Delta U)$ taken to
be zero. With this setup
the systematic error was estimated to be less
than $0.02\,\mbox{kJ}\,\mbox{mol}^{-1}$. Although the construction of the
excluded volume list needs a little
additional computational effort, 
this simple scheme improves the efficiency of the sampling       
by almost two orders of magnitude.
For the calculation of the Lennard-Jones insertion energies
$\Delta U$ we have used cut-off distances of $9\,\mbox{\AA}$ 
in combination with a proper cut-off correction. Each configuration
has been probed by $10^3$ {\em successful} insertions (i.e. insertions into the
free volume, contributing non-vanishing Boltzmann-factors).

We use simulations containing 500 Water molecules and 8 Xenon particles to
study the temperature dependence of the association behavior of Xenon.
The hydrophobic interaction between the dissolved Xenon particles
is quantified in terms the profile of free energy (PMF) for the association of
two Xenon particles.
The $w(r)$ is obtained by inverting the Xenon-Xenon radial distribution
functions $g(r)$
according to
\begin{equation}
w(r)=-kT \ln g(r)\;.
\end{equation}
We use temperature derivatives of quadratic fits of $w(r,T)$ 
to calculate
the enthalpic and entropic contributions at each Xenon-Xenon separation
$r$. 
All five temperatures 
$275\,\mbox{K}$, $300\,\mbox{K}$, $325\,\mbox{K}$, $350\,\mbox{K}$
and $375\,\mbox{K}$ were taken into account for the fits .
The entropy and enthalpy contributions are then obtained as 
\begin{equation}
s(r)=-\left(\frac{\partial w(r,T)}{\partial T}\right)_P
\end{equation}
and
\begin{equation}
h(r)=w(r)+T s(r) \;.
\end{equation}
In addition, the corresponding heat capacity change relative to the bulk 
liquid is available according to
\begin{equation}
c_P(r) = -T \left( \frac{\partial^2 w(r,T)}{\partial^2 T} \right)_P\;.
\end{equation}

\subsection{``Calorimetric'' Analysis}

\label{sec:calory}

In order to provide a spacial resolution of 
the water contribution to the solvation excess
heat capacity, we first calculate the individual 
potential energies of
the water and solute molecules. This is done
by a reaction field method
based on the minimum image convention in combination with 
a minimum image ``cubic'' cutoff. This approach has been
originally proposed by Neumann \cite{Neumann:83} 
and was discussed by Roberts and
Schnitker \cite{Roberts:94,Roberts:95}.
The reaction field approach in general is well suited
for our purposes since it
makes it easy to cleanly according the potential 
energy contributions to individual molecules.
For convenience we divide the potential energies in contributions
assigned to the individual molecules with
\begin{eqnarray}
  E
  &=&
  \sum_{i=1}^M E_i \nonumber \\
  E_i
  &=&
  \left(\frac{1}{2}\sum_{j=1}^M E_{ij}\right) + E_{i,corr.} \;,
\end{eqnarray}
where $E_i$ is the potential energy assigned to molecule $i$,
$M$ is the total number of molecules. The molecule-molecule
pair energy
\begin{eqnarray}
  E_{ij}
  &=&
  \sum_\alpha \sum_\beta 
  4\,\epsilon_{i\alpha j\beta}
  \left[
    \left(
      \frac{\sigma_{i\alpha j\beta}}{r_{i\alpha j\beta}}
    \right)^{12}
    -
    \left(
      \frac{\sigma_{i\alpha j\beta}}{r_{i\alpha j\beta}}
    \right)^{6}
  \right]
  \nonumber\\
 & &
  \hspace*{3em}+ \frac{q_{i\alpha}q_{j\beta}}{r_{i\alpha j\beta}}
\end{eqnarray}
is then obtained as sum over discrete interaction sites $\alpha$
and $\beta$, with $r_{i\alpha j\beta}\!=\!|\vec{r}_{j\beta}-\vec{r}_{i\alpha}|$
based on the molecule/molecule center of mass minimum image separation.
We employ long range corrections
$E_{i,corr.}\!=\!E^{el}_{i,corr.}+E^{LJ}_{i,corr.}$
accounting for electrostatic, as well as Lennard
Jones interactions. The electrostatic correction 
\begin{eqnarray}
  E^{el}_{i,corr.}
  &=&
  \frac{2\pi}{3 V}\vec{D}\,\vec{d}_i
\end{eqnarray} 
is a reaction field term, corresponding to the
cubic cutoff, assuming an infinitely large dielectric
dielectric constant. Here 
$\vec{d}_i\!=\!\sum_\alpha q_{i\alpha}\vec{r}_{i\alpha}$
is the dipole moment of
molecule $i$, $\vec{D}=\sum_i \vec{d}_i$ 
is the total dipole moment of all molecules in the simulation cell
and $V$ is the instantaneous volume of the simulation box.
$E^{el}_{i,corr.}$ has also been considered as the {\em extrinsic} potential
and has been shown to provide configurational energies 
quite close to the values 
obtained by Ewald summation (with tin-foil boundary conditions)
\cite{Roberts:94}.
In order to be consistent with applied cubic cutoff procedure for
the electrostatic interactions,
we also use a Lennard Jones correction term for the cubic cutoff
\begin{eqnarray}
  E^{LJ}_{i,corr.}
  &=&
  \frac{2}{V}
  \sum_\alpha
  \sum_j\sum_\beta
  \frac{\kappa_6}{b^3}
  \left(-\epsilon_{i\alpha j\beta}\sigma_{i\alpha j\beta}^6\right)
  \nonumber\\
  & & \hspace*{1em}
  +\frac{\kappa_{12}}{b^9}
  \left(\epsilon_{i\alpha j\beta}\sigma_{i\alpha j\beta}^{12}\right)\;,
\end{eqnarray}
with $b\!=\!V^{1/3}/2$ denoting the half box length.
$\kappa_6\!=\![2+15\sqrt{2} \arctan(1/\sqrt{2})]/6\approx2.5093827$
and
$\kappa_{12}\!=\![17774+77409\sqrt{2} 
\arctan(1/\sqrt{2})]/207360\approx0.4106497$
are analytically integrated factors accounting for the 
cubic cutoff geometry. In case of the more common spherical cutoff, one would
have to replace $\kappa_6/b^3$ by $4\pi/3 R_c^3$ and
$\kappa_{12}/b^9$ by $4\pi/9 R_c^9$, with $R_c$ being the cutoff-radius.

In addition to the procedure outlined 
section \ref{sec:hydrophobic}, we can as well directly use the 
individual energies to
 calculate the 
the solvation enthalpies and heat capacities 
according to
\begin{eqnarray}
  h_{ex}
  &=&
  \left<E_{\rm solute}\right> + \nonumber \\
  & &
  \left<n_{\rm shell}\right> \times
  \left(
    \left< E_{\rm shell}\right>-\left< E_{\rm bulk}\right>
  \right)
\end{eqnarray}
where $\left<E_{\rm solute}\right>$ is the average potential energy of
the solute molecule, $\left< E_{\rm shell}\right>$ is
the average potential energy of the water molecules in a 
sufficiently large solvation sphere (here we use a radius of $1.0\,\mbox{nm}$)
around the solute molecule, whereas $\left< E_{\rm bulk}\right>$ is
the energy of the water molecules outside this sphere.
$\left<n_{\rm shell}\right>$ is the corresponding number of water molecules
in the solvation sphere. From the temperature dependence of $h_{ex}$
we can obtain the corresponding heat capacities straightforwardly.


\section{DISCUSSION}

\subsection{Hydrophobic Hydration}
\label{sec:hydration}

The hydration free energies $\mu_{ex}$ for Xenon,
given in Table \ref{tab:muex},
were obtained for a system of 500 SPCE 
molecules employing
the Widom particle insertion method. In Figure \ref{fig:n01} these
data are shown, as well as a quadratic fit of the 
data with respect to the temperature.
The corresponding solvation entropies and enthalpies are derived
from the fitted temperature dependence, and are depicted as dotted and
dashed lines in Figure \ref{fig:n01}, respectively.
A detailed comparison of the thermodynamic solvation properties 
of Xenon and other noble gases using different water models, with
experimental data, however, has been the 
subject of a previous publication \cite{Paschek:2004}. In this contribution we
would like to focus on the energetical and structural
details regarding the hydration and association of
hydrophobic Xenon particles. 

\ifpretty
\begin{table}[!b]
  \centering
  \ifpretty
  \renewcommand{\tabcolsep}{1.5cm}
  \renewcommand{\arraystretch}{1.0}
  \else
  \renewcommand{\tabcolsep}{3.3cm}
  \renewcommand{\arraystretch}{1.0}
  \fi
  \small
  \begin{tabular}{cc} \\ \hline\hline \\[-4pt]
  $T/\mbox{K}$ &
  $\mu_{ex}/\mbox{kJ}\,\mbox{mol}^{-1}$ 
\\[6pt] \hline \\[-6pt]
$275$ & $6.15\pm0.1$ \\
$300$ & $7.71\pm0.1$ \\
$325$ & $9.05\pm0.1$ \\
$350$ & $9.87\pm0.1$ \\
$375$ & $10.41\pm0.1$ 
\\[6pt] \hline\hline
  \end{tabular}
  \caption{\footnotesize
    Calculated excess chemical potential $\mu_{ex}$ for Xenon dissolved in
    SPCE water, 
    obtained from the 500 molecule systems by the Widom particle 
    insertion technique.}
  \label{tab:muex}
\end{table}

\begin{table*}[!t]
  \centering
  \ifpretty
  \renewcommand{\tabcolsep}{0.85cm}
  \renewcommand{\arraystretch}{1.0}
  \else
  \renewcommand{\tabcolsep}{0.4cm}
  \renewcommand{\arraystretch}{1.0}
  \fi
  \small
  \begin{tabular}{ccc|c} \\ \hline\hline \\[-4pt]
  ~& (500 SPCE) & (500 SPCE + 1 Xe) 
  \\[2pt]
  $T/\mbox{K}$ &
  $\left< E_{c}\right>/\mbox{kJ}\,\mbox{mol}^{-1}$ &
  $\left< E_{c}\right>/\mbox{kJ}\,\mbox{mol}^{-1}$  & 
  $\left< \Delta E_{c}\right>/\mbox{kJ}\,\mbox{mol}^{-1}$ 
\\[6pt] \hline \\[-6pt]
$275$ & $-24073.6\pm1.3$\hspace{1em}($-48.1472 \pm 0.0026$)  &  $-24086.6\pm1.0$   &  $-13.0\pm1.6$ \\
$300$ & $-23288.1\pm0.7$\hspace{1em}($-46.5762 \pm 0.0014$)  &  $-23296.8\pm0.7$   &  $-8.8\pm1.0$  \\
$325$ & $-22531.2\pm0.5$\hspace{1em}($-45.0624 \pm 0.0010$)  &  $-22536.2\pm0.8$   &  $-5.0\pm0.9$  \\
$350$ & $-21794.3\pm0.7$\hspace{1em}($-43.5886 \pm 0.0014$)  &  $-21796.0\pm0.6$   &  $-1.7\pm0.9$  \\
$375$ & $-21071.2\pm0.8$\hspace{1em}($-42.1424 \pm 0.0016$)  &  $-21068.7\pm0.8$   &  $ 2.5\pm1.1$  
\\[6pt] \hline\hline
  \end{tabular}
  \caption{\footnotesize Left columns:
    Average configurational energies $E_c$ as directly obtained from the simulations
    of 500 SPCE molecules with and without an additional Xenon particle. 
    The values given in brackets denote the energy per molecule. The data
    corresponds to
    the Particle mesh Ewald summation, applying appropriate cutoff
    corrections for the Lennard-Jones contributions. Right column: The difference
    between the two energy contributions denotes the solvation enthalpy 
    $h_{ex}\!\equiv\!\left< \Delta E_c\right>$ for
    Xenon in SPCE water.}
  \label{tab:econf}
\end{table*}

\begin{table*}[!t]
  \centering
  \ifpretty
  \renewcommand{\tabcolsep}{0.46cm}
  \renewcommand{\arraystretch}{1.0}
  \else
  \renewcommand{\tabcolsep}{0.15cm}
  \renewcommand{\arraystretch}{1.0}
  \fi
  \small
  \begin{tabular}{ccccc|c} \\ \hline\hline \\[-4pt]
  $T/\mbox{K}$ &
  $\left< E_{\rm solute}\right>/\mbox{kJ}\,\mbox{mol}^{-1}$ &
  $\left< E_{\rm shell}\right>/\mbox{kJ}\,\mbox{mol}^{-1}$  & 
  $\left< n_{\rm shell}\right>$  & 
  $\left< E_{\rm bulk}\right>/\mbox{kJ}\,\mbox{mol}^{-1}$  & 
  $\left< \Delta E\right>/\mbox{kJ}\,\mbox{mol}^{-1}$ 
\\[6pt] \hline \\[-6pt]
$275$ & $-10.196\pm0.014$  & $-48.125\pm0.016$ & $137.50\pm0.04$ &  $-48.102\pm0.007$ & $-13.5\pm2.4$\\
$300$ & $ -9.846\pm0.014$  & $-46.527\pm0.014$ & $135.94\pm0.03$ &  $-46.527\pm0.006$ & $ -9.9\pm2.0$\\
$325$ & $ -9.453\pm0.013$  & $-44.980\pm0.013$ & $133.97\pm0.03$ &  $-45.020\pm0.005$ & $ -4.1\pm1.9$\\
$350$ & $ -9.071\pm0.012$  & $-43.489\pm0.016$ & $131.28\pm0.04$ &  $-43.545\pm0.005$ & $ -1.8\pm2.2$\\
$375$ & $ -8.672\pm0.014$  & $-42.017\pm0.014$ & $128.18\pm0.03$ &  $-42.099\pm0.005$ & $  1.9\pm1.9$
\\[6pt] \hline\hline
  \end{tabular}
  \caption{\footnotesize
    Potential energies of the solute particle and the water molecules 
    obtained by using a cubic minumum image cutoff, a reaction field
    correction and a cubic
    Lennard-Jones cutoff correction, as outlined in the text.
    $\left< E_{\rm solute}\right>$ is the potential energy of the solute particle,
    $\left< E_{\rm shell}\right>$ is the average potential energy of the water molecules in the spehere
    of $1.0\,\mbox{nm}$ radius around the solute particle.
    $\left< n_{\rm shell}\right>$ is the average number of water molecule in this sphere.
    $\left< E_{\rm bulk}\right>$ is the potential energy of the water molecules
    outside the solute
    sphere. The net solvation enthalpy is obtained as 
    $h_{ex}\equiv \left< \Delta E\right>\!=\!\left< E_{\rm solute}\right>+
    \left< n_{\rm shell}\right>\times(\left< E_{\rm shell}\right>-\left<
    E_{\rm bulk}\right>)$.
    }
  \label{tab:edetail}
\end{table*}

\begin{figure}[!t]
  \centering
  \includegraphics[angle=0,width=7.0cm]{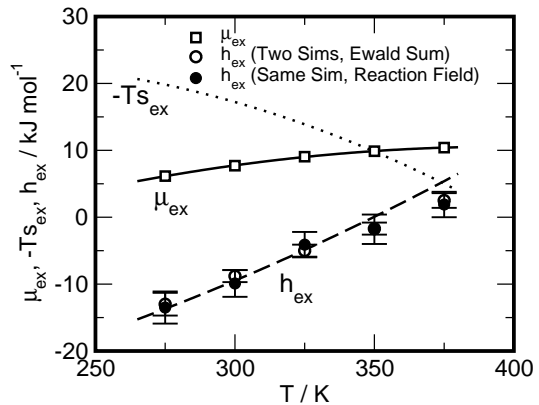}
  \caption{\footnotesize Squares: Excess chemical potential of Xenon
   in SPCE water. Solid line: Cubic fit to the excess chemical potential
   of Xenon with respect to the temperature according to
   $\mu_{ex}(T)\!=\!\mu_0 + \mu_1 T + \mu_2 T^2$ with
   $\mu_0\!=\!-35.90\,\mbox{kJ}\,\mbox{mol}^{-1}$,
   $\mu_1\!=\!2.33\times 10^{-1}\,\mbox{kJ}\,\mbox{mol}^{-1}\,\mbox{K}^{-1}$
   and
   $\mu_2\!=\!-2.94\times 10^{-4}\,\mbox{kJ}\,\mbox{mol}^{-1}\,\mbox{K}^{-2}$.
   Dotted line: Solvation entropy $s_{ex}$
   according to the fitted data.
   Dashed line: Solvation enthalpy $h_{ex}$
   according to the fitted data.
   Open circles: Solvation enthalpies obtained directly by subtracting the
   total configurational energies from simulations consisting of 500 SPCE 
   + 1 Xenon molecules and simulations containing 500 SPCE molecules.
   The energy data are
   according to the particle mesh Ewald summation.
   Filled circles: Solvation enthalpies obtained from
   500 SPCE + 1 Xenon simulations by using the individually calculated
   energies for the solute, shell water and bulk water. The energies
   are obtained by using the reaction field method discussed in the
   text.
  }
  \label{fig:n01}
\end{figure}

\begin{figure}[!t]
  \centering
  \includegraphics[angle=0,width=7.0cm]{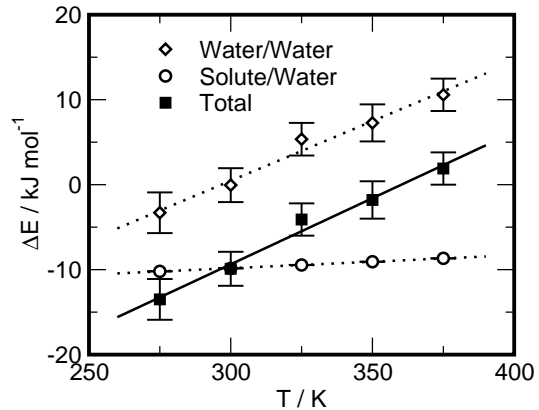}
  \caption{\footnotesize 
    Solvation enthalpies $\Delta E\equiv h_{ex}$ obtained from the
    500 SPCE + 1 Xenon simulations by using the individually calculated
    energies for the solute, shell water and bulk water:
    $\Delta E_{\rm Solute/Water}\!=\!E_{\rm solute}$ and
    $\Delta E_{\rm Water/Water}\!=\!n_{\rm shell}\times(E_{\rm shell}-E_{\rm bulk})$.
    The open diamonds denote the 
    Water/Water contribution, whereas the the open circles indicate
    the Solute/Water contribution.
    The black squares indicate the sum of both contributions.
    The lines indicate the excess heat capacities:
    Solute/Water:
    $c_{P,ex}\!=\!15.3\,\mbox{J}\,\mbox{K}^{-1}\,\mbox{mol}^{-1}$;
    Water/Water:
    $c_{P,ex}\!=\!140.3\,\mbox{J}\,\mbox{K}^{-1}\,\mbox{mol}^{-1}$;
    Total
    $c_{P,ex}\!=\!155.6\,\mbox{J}\,\mbox{K}^{-1}\,\mbox{mol}^{-1}$.
    }
  \label{fig:n02}
\end{figure}

\fi
In addition to the indirect determination of the solvation
enthalpy based on the temperature dependence
of the solvation free energy, also a direct calculation is possible.
Therefore we have to consider
the potential energy of the solute molecule, as well as the change of
the energy of the water molecules in the solvation sphere relative to
the bulk. Another {\em brute force}
approach according to Durell and Wallqvist \cite{Durell:96}, 
however, is to determine the energy difference between 
simulations of two distinct systems, one containing only the solvent water,
and another, consisting of water plus an additional solute molecule.
This procedure has also been applied to the case of Xenon and SPCE water.
The configurational energies shown in Table \ref{tab:econf}
were  directly obtained from GROMACS during the simulation, employing
the particle mesh Ewald summation technique (see section \ref{sec:MD} for
details concerning the setup of the PME). 
The rather long simulation runs of $20\,\mbox{ns}$ length
provide us sufficiently accurate data to determine the hydration
enthalpies with an errorbar of approximately
$\pm1\,\mbox{kJ}\,\mbox{mol}^{-1}$.
A comparison of the directly obtained $h_{ex}$ data with the hydration enthalpies
determined from the solvation free-energies is given in 
Figure \ref{fig:n01}. The dashed curve, which is representing
the hydration enthalpies according to the fitted $\mu_{ex}(T)$ data,
is mostly lying within the errorbars of the 
directly obtained enthalpies (open
symbols). However, the apparent deviation at $375\,\mbox{K}$ might be attributed
to the restriction of the fit of the $\mu_{ex}$-data 
to the finite temperature interval between $275\,\mbox{K}$
and $375\,\mbox{K}$.

\ifpretty
\begin{figure}[!t]
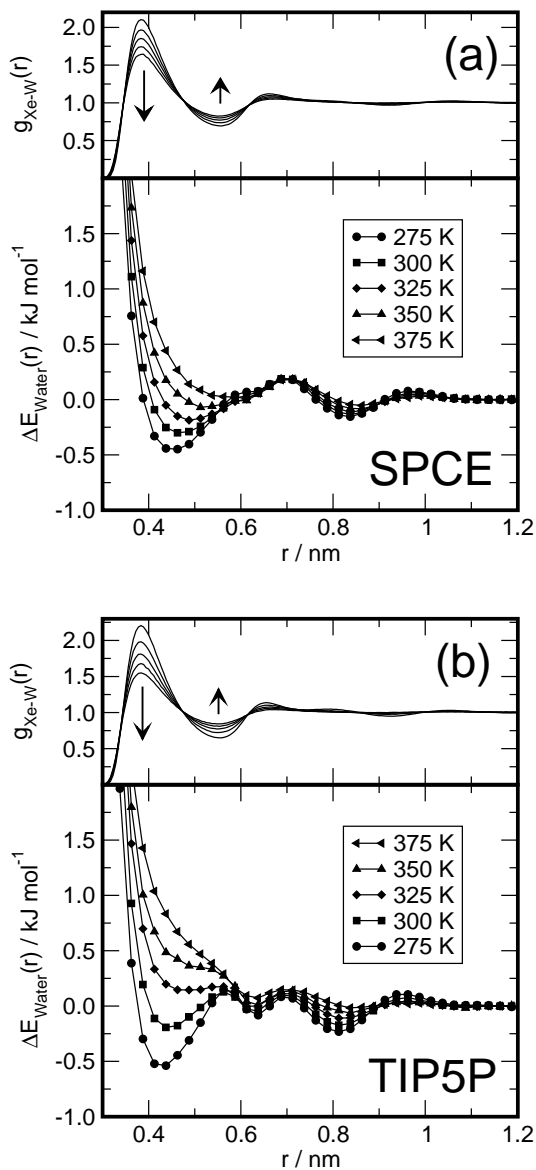

  \centering
  \includegraphics[angle=0,width=7.0cm]{FIG03a}

  \vspace*{2em}
  \includegraphics[angle=0,width=7.0cm]{FIG03b}
  \caption{\footnotesize 
    Xenon-Water center of mass pair distribution 
    function $g_{\rm Xe-W}(r)$
    and relative change of the water potential energy
    $\Delta E_{\rm Water}(r)\!=\! E_{\rm Water}(r)- E_{\rm Water}^{bulk}$ for
    all investigated temperatures. 
    The arrows indicate the sequence of the
    $g(r)$-curves pointing
    from low to high temperatures.
    a. SPCE model b. TIP5P model.
    }
  \label{fig:n03}
\end{figure}

\begin{figure}[!t]
  \centering
  \includegraphics[angle=0,width=7.0cm]{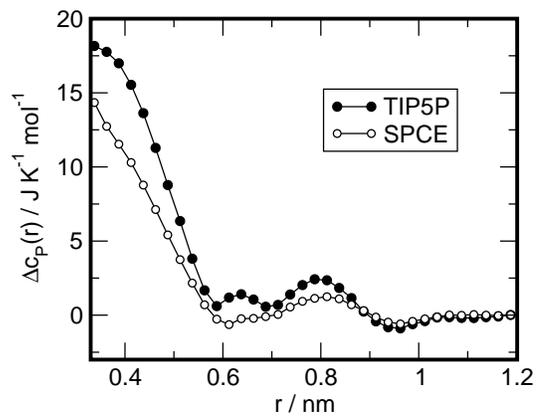}
  \caption{\footnotesize
    Change of the heat capacity of the water molecules around a Xenon particle
    $\Delta c_P(r)\!=\! c_P(r)- c_P^{bulk}$ 
    as a function of distance to the Xenon particle.
    The $\Delta c_P(r)$ is obtained as a linear regression of the datasets shown in
    Figure \ref{fig:n03}.
    }
  \label{fig:n04}
\end{figure}

\fi
As an alternative to the {\em two simulations} approach, we calculate
the solvation enthalpy based on the ``individual'' potential energies of
solute and solvent molecules using the reaction field method discussed in
section \ref{sec:calory}. Table \ref{tab:edetail} contains the energies of the
solute molecule $E_{\rm solute}$ and the water molecules in the bulk and in
the hydration sphere of radius $1.0\,\mbox{nm}$, as well as the
number of the water molecules in the hydration sphere. A 
comparison of the energy values for the water bulk using the
reaction field with the data for the pure water system according to 
the Ewald summation (given in Table \ref{tab:econf}) reveals that
the reaction field data are lying systematically about $0.042\,\mbox{kJ}\,\mbox{mol}^{-1}$
higher than the Ewald data. Although being rather small, this noticeable
difference is of systematic nature and
is presumably due to the lack of higher order
multipole contributions to the reaction field term, which are, of course, present
in the Ewald calculation \cite{Boresch:2001}.
However, the differences apparently cancel out when considering 
energy differences between shell and bulk, when determining the solvation
enthalpies. As shown in Figure \ref{fig:n01} and Tables
\ref{tab:econf} and \ref{tab:edetail}, the obtained data for $h_{ex}$ is, both, consistent
with the {\em two simulations} solvation enthalpies, as well as the solvation
enthalpies obtained from the temperature dependence of the solvation free energies.
Figure \ref{fig:n02} reports a division of the solvation enthalpy
$h_{ex}$ (total) into contributions according to the solute 
$\left< E_{\rm solute}\right>$ (Solute/Water) and the solvent
$h_{ex}-\left< E_{\rm solute}\right>$ (Water/Water). The 
calculated total excess heat
capacity $c_{P,ex}$ of $155.6\,\mbox{J}\,\mbox{K}^{-1}\,\mbox{mol}^{-1}$
is dominated by the Water/Water contribution of
$140.3\,\mbox{J}\,\mbox{K}^{-1}\,\mbox{mol}^{-1}$,
whereas the Solute/Water part with
$15.3\,\mbox{J}\,\mbox{K}^{-1}\,\mbox{mol}^{-1}$
contributes only to about 10\%. 
The observed value for $c_{P,ex}$ of 
$155.6\,\mbox{J}\,\mbox{K}^{-1}\,\mbox{mol}^{-1}$, 
however, does only qualitatively 
agree with the value of about $280\,\mbox{J}\,\mbox{K}^{-1}\,\mbox{mol}^{-1}$, 
observed experimentally for $T\!=\!300\,\mbox{K}$
\cite{Paschek:2004,Prini:89}. We would also like to point out that the 
experimentally observed increase of the excess heat capacity with 
decreasing temperature (as discussed in Ref. \cite{Southall:2002}) 
seems not to be present in the SPCE simulation data.
However, the change of the Xenon solvation enthalpy
from negative to positive at about $363\,\mbox{K}$ has to be attributed mainly
to the potential energy change of the of water molecules in the 
hydrophobic hydration shell.
In order to elucidate the spacial (radial) extension of the solvent contribution
to the excess heat capacity, we calculate the potential
energy of the water molecules as a function of distance to the Xenon atom.
Figure \ref{fig:n03} shows the change of the potential energy
of the water molecules around a Xenon particle with respect to the bulk value
for the SPCE and TIP5P models as a function of temperature. 
Due to the better statistics, the data
in Figure \ref{fig:n03} were obtained from the simulations containing 8
Xenon particles. 
For completeness,
also the Xenon-Water center of mass pair distribution functions are 
indicated. A rather strong
temperature dependence of the potential energy of the water molecules in the
first hydration shell is clearly evident for both models. 
A rather remarkable observation is, that for 
the lower temperatures the water molecules
in the distance interval between $0.4\,\mbox{nm}$ and $0.5\,\mbox{nm}$,
which are corresponding to the first hydration shell,
exhibit a potential energy 
even lower than the bulk value. With increasing temperature
this behavior is reversed and the location of the molecules in the hydration
shell becomes more and more energetically unfavorable. On a qualitative level, both
the SPCE and TIP5P model, exhibit a similar behavior. The TIP5P model, however,
shows a more strongly pronounced temperature dependence, and a more richly 
structured potential energy profile. 
We would like to emphasize  that for the SPCE model at the
 lowest temperature we find  
an average potential energy for the water molecules in the hydration
sphere, which lies {\em below} the average bulk value 
(see Table \ref{tab:edetail}). The fact that
Durell and Wallqvist did not observe this in their simulations
\cite{Durell:96} is perhaps due to their restricted
temperature interval ($300\,\mbox{K}$ to $350\,\mbox{K}$).

The temperature dependence
of the distance dependent water potential 
energies are quantified in terms of the configurational heat capacity.
Figure \ref{fig:n04} shows the water heat capacities
as a function of the distance to the Xenon particle.
The data are given relative to the value for the water bulk of
$60.0\,\mbox{J}\,\mbox{K}^{-1}\,\mbox{mol}^{-1}$,
obtained for the SPCE model,
and $90.8\,\mbox{J}\,\mbox{K}^{-1}\,\mbox{mol}^{-1}$,
obtained for the TIP5P model.
As already evident from Figure \ref{fig:n03}, the strongest heat capacity
effect is observed for the first hydration shell.
For the TIP5P model a noticeably enhanced heat capacity 
is also observed for the distance range between 
$0.6\,\mbox{nm}$  and $1.0\,\mbox{nm}$.
Beyond a separation distance of 
$1.0\,\mbox{nm}$, however, no significant energy difference 
compared to the bulk can be denoted, indicating a
properly chosen size of the solvation sphere used for the calculation
of the hydration enthalpies $h_{ex}$.

\subsection{Hydrophobic Interaction}

\label{sec:interaction}

\ifpretty
\begin{figure}[!b]
  \centering
  \includegraphics[angle=0,width=7.0cm]{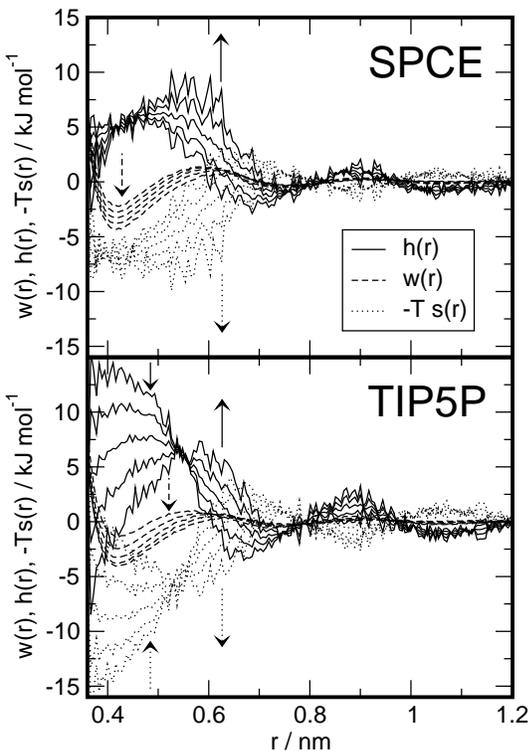}
  \caption{\footnotesize
    Profiles of free energy $w(r)$ obtained
    for the association of two Xenon particles (dashed lines)
    as well as their enthalpic and entropic contributions
    $w(r)=h(r)-Ts(r)$ for all five temperatures 
    ($275\,\mbox{K}$, 
    $300\,\mbox{K}$, 
    $325\,\mbox{K}$, 
    $350\,\mbox{K}$, 
    $375\,\mbox{K}$).
    Solid lines: $h(r)$. Dotted lines: $-Ts(r)$
    Top: SPCE model; Bottom:
    TIP5P model. The arrows indicate the sequence of the
    data curves pointing
    from low to high temperatures.
    The data shown here is identical to the
    data presented in Ref. \cite{Paschek:2004}.
    }
  \label{fig:n05}
\end{figure}

\begin{figure}[!t]
  \centering
  \includegraphics[angle=0,width=7.0cm]{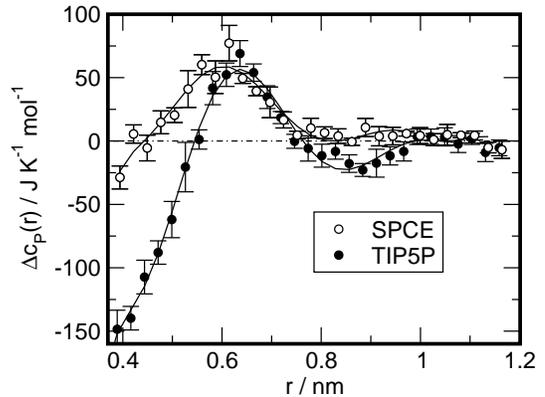}
  \caption{\footnotesize
    Relative change in heat capcacity 
    $\Delta c_P(r)\!=\! c_P(r)- c_P^{bulk}$
    for the hydrophobic interaction
    between two dissolved Xenon particles 
    using the SPCE and TIP5P models.
    The data correspond to the temperature derivative of quadratic fits
    of $w(r,T)$  obtained for $300\,\mbox{K}$. The solid lines are just to 
    guide the eye. The data shown here is identical to the data
    reported in Ref. \cite{Paschek:2004}.
    }
  \label{fig:n06}
\end{figure}

\begin{figure}[!t]
  \centering
  \includegraphics[angle=0,width=4.0cm]{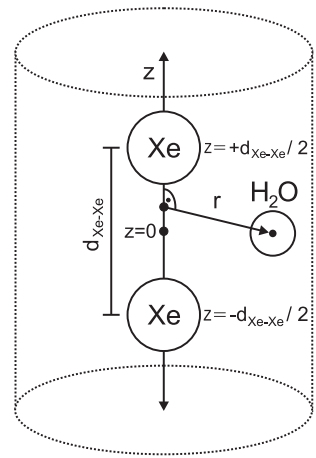}
  \caption{\footnotesize
    Schematic illustration of the employed definition for the
    cylindrical distribution of the water density
    and heat capacity around a pair of Xenon particles
    obtained for different Xenon-Xenon separation distances 
    $d_{\rm Xe-Xe}$.
    }
  \label{fig:n07}
\end{figure}

\begin{figure*}[!t]
  \centering
  \includegraphics[angle=0,width=12cm]{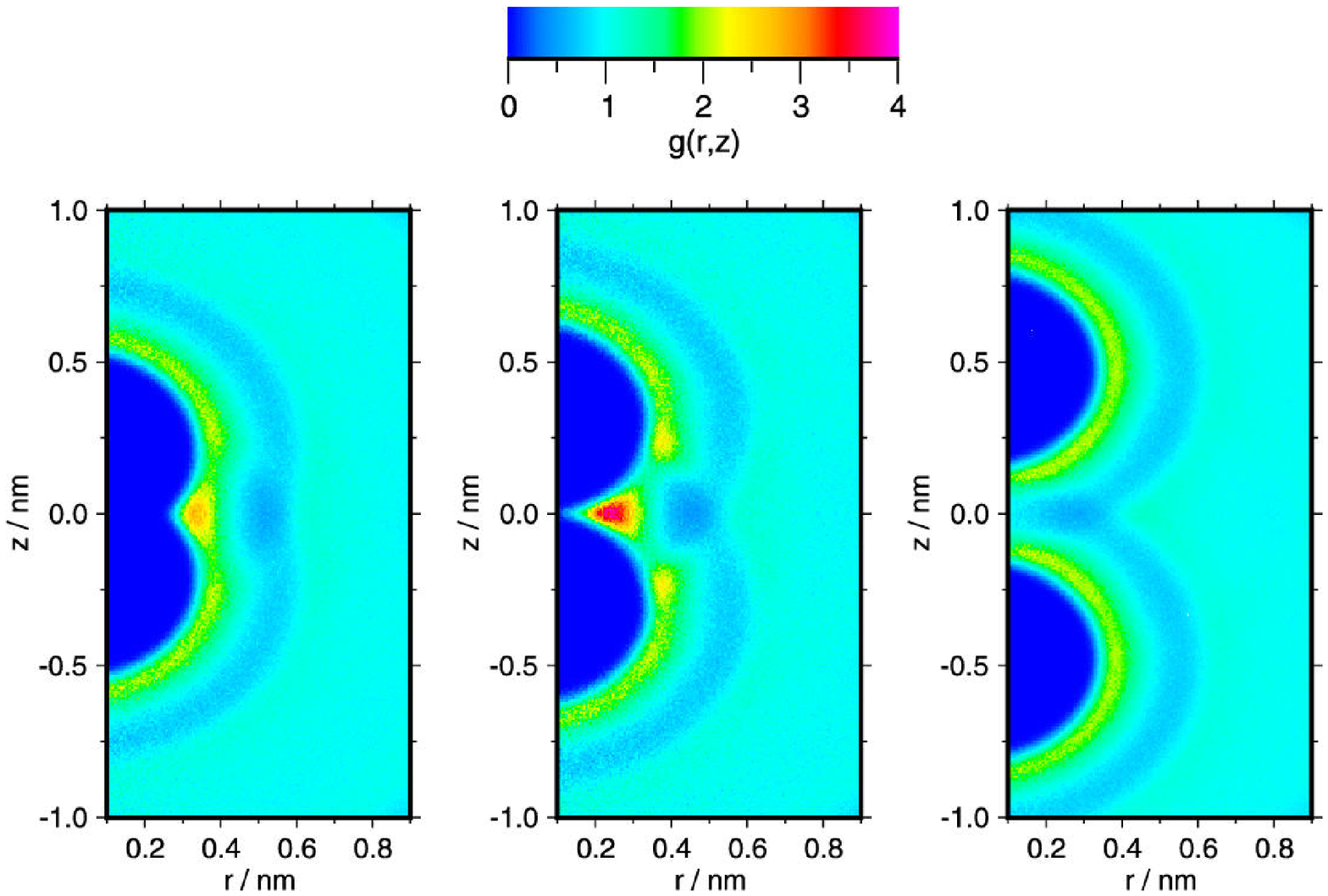}
  \caption{\footnotesize
    Cylindrical distribution function of the SPCE water molecules center of mass
    around a pair of Xenon particles 
    $g_{\rm Xe-Xe-W}(r,z)$ obtained for a certain Xe-Xe 
    distance interval at $T\!=\!300\,\mbox{K}$. 
    Left: ``contact'' ($d_{\rm Xe-Xe}\!\leq\!0.45\,\mbox{nm}$).
    Middle: ``desolvation barrier'' 
    ($0.55\,\mbox{nm}\!\le\!d_{\rm Xe-Xe}\!\leq\!0.65\,\mbox{nm}$).
    Right: ``separated''
    ($0.9\,\mbox{nm}\!\le\!d_{\rm Xe-Xe}\!\leq\!1.0\,\mbox{nm}$).
    }
  \label{fig:n08}
  \vspace*{1em}
  \includegraphics[angle=0,width=12cm]{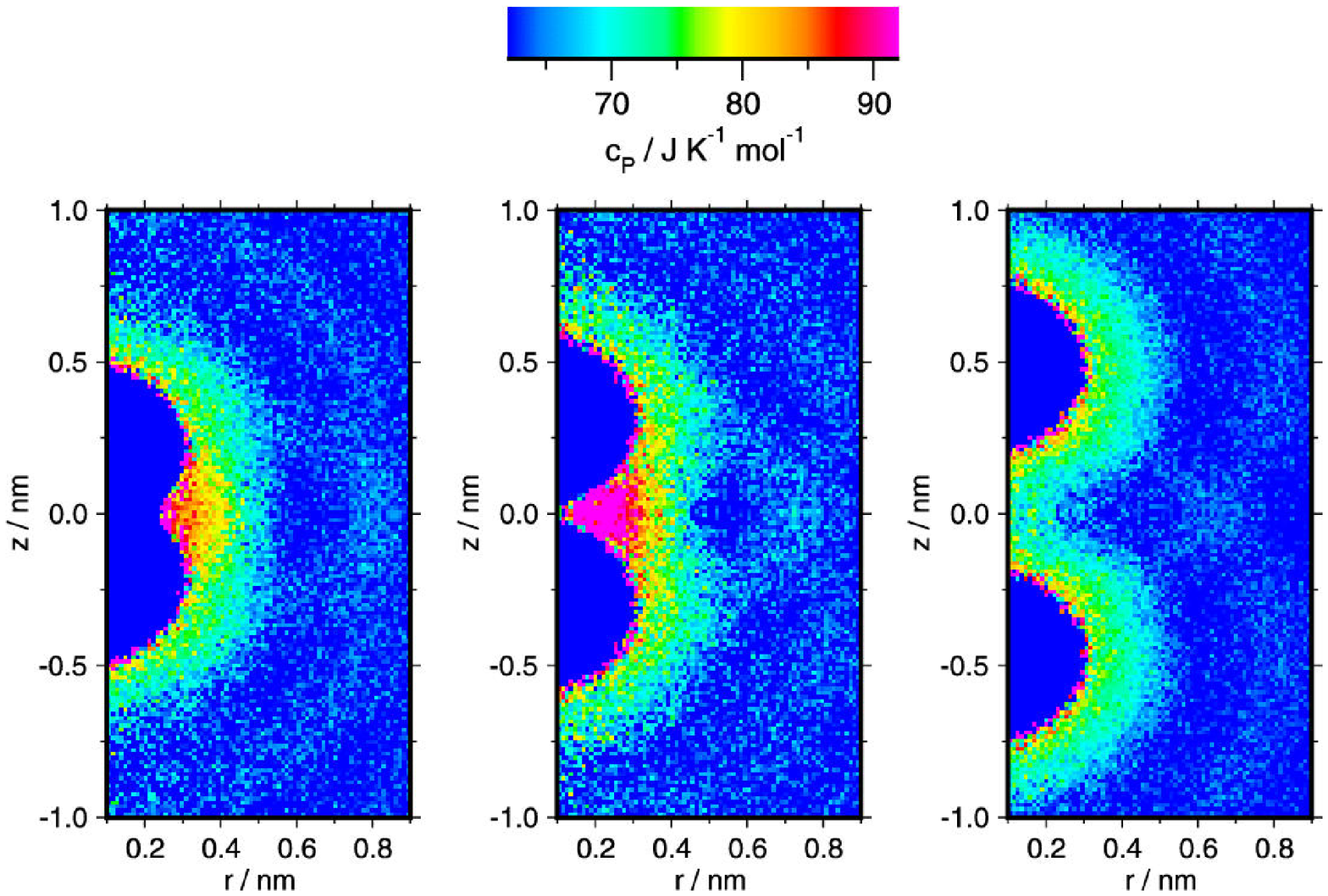}
  \caption{\footnotesize
    Cylindrical distribution function of the configurational contribution to the
    heat capacity of the SPCE water molecules
    around a pair of Xenon particles $c_P(r,z)$ obtained for a certain Xe-Xe 
    distance interval. The $c_P(r,z)$ data have been obtained as 
    linear fits to the corresponding water potential energies $E_{Water}(r,z)$
    for all five temperatures.
    The average bulk value is $62.1\,\mbox{J}\,\mbox{K}^{-1}\mbox{mol}^{-1}$.
    Left: ``contact'' ($d_{\rm Xe-Xe}\!\leq\!0.45\,\mbox{nm}$).
    Middle: ``desolvation barrier'' 
    ($0.55\,\mbox{nm}\!\le\!d_{\rm Xe-Xe}\!\leq\!0.65\,\mbox{nm}$).
    Right: ``separated''
    ($0.9\,\mbox{nm}\!\le\!d_{\rm Xe-Xe}\!\leq\!1.0\,\mbox{nm}$).
}
  \label{fig:n09}
\end{figure*}


\fi
The hydrophobic interaction between two Xenon particles 
is obtained as a profile of free energy $w(r)$ for the association
of Xenon particles from the simulations containing 8 Xenon particles.
The $w(r)$, as well as the corresponding enthalpic and
entropic contributions $h(r)$ and $-Ts(r)$
are shown in Figure \ref{fig:n05}.
For both water models the ``contact state'' is defined by the  minimum  of
the $w(r)$ function at a distance of about $0.42\,\mbox{nm}$.
With increasing temperature the well depth of the minimum of the profile of free energy
at the contact state drops
from about $-2.21\,\mbox{kJ}\,\mbox{mol}^{-1}$ ($275\,\mbox{K}$) to
$-4.25\,\mbox{kJ}\,\mbox{mol}^{-1}$ ($375\,\mbox{K}$)
for the SPCE model and $-1.36\,\mbox{kJ}\,\mbox{mol}^{-1}$ ($275\,\mbox{K}$) to
$-4.02\,\mbox{kJ}\,\mbox{mol}^{-1}$ ($375\,\mbox{K}$) for the TIP5P model. The
accuracy for each of the $w(r)$ profiles has been determined to be about
$\pm0.15\,\mbox{kJ}\,\mbox{mol}^{-1}$.
The contact state minimum and the minimum characterizing the
solvent separated state, which is located at
a distance of about $0.72\,\mbox{nm}$ to $0.78\,\mbox{nm}$,
are separated by the so called {\em desolvation barrier}, which is
found approximately at a distance of $0.6\,\mbox{nm}$. 
Figure \ref{fig:n05} indicates that for lower temperatures the contact state is entropically
stabilized and enthalpically destabilized, in accordance
with the observation of Smith and Haymet \cite{Smith.D:92,Smith.D:93}
and others \cite{Luedemann:96,Rick:97,Shimizu:2000,Rick:2000,Ghosh:2002}.
For the TIP5P model, we observe
a strong temperature dependence of the curves, 
showing a decrease of the absolute values
for the enthalpy and entropy contributions. In contrast to the TIP5P model,
the SPCE model does exhibit 
only a weak temperature dependence of $h(r)$ and $-Ts(r)$
at the contact state .
At the desolvation barrier, however, the inverse behavior is observed
for both models: with
increasing temperature the entropy/enthalpy compensation effect is found to
be enlarged. 

\ifpretty
\begin{figure}[!t]
  \centering
  \includegraphics[angle=0,width=7.0cm]{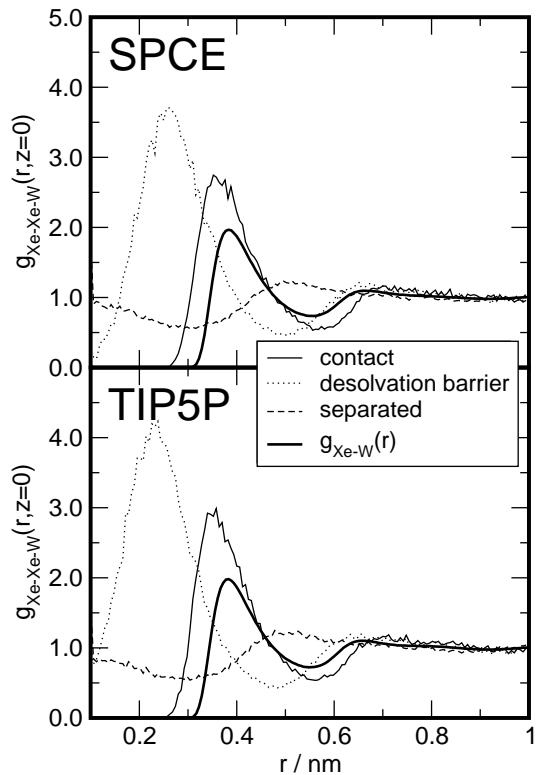}
  \caption{\footnotesize
    Distribution function of the SPCE water molecules center of mass
    around a pair of Xenon particles in the Xe-Xe bisector plane
    $g_{\rm Xe-Xe-W}(r,z\!=\!0)$ obtained for a certain Xe-Xe 
    distance interval at $T\!=\!300\,\mbox{K}$. 
    ``contact'': $d_{\rm Xe-Xe}\!\leq\!0.45\,\mbox{nm}$.
    ``desolvation barrier'':
    $0.55\,\mbox{nm}\!\le\!d_{\rm Xe-Xe}\!\leq\!0.65\,\mbox{nm}$.
    ``separated'':
    $0.9\,\mbox{nm}\!\le\!d_{\rm Xe-Xe}\!\leq\!1.0\,\mbox{nm}$.
    For comparison, also the Xe-W pair distribution functions are given.
    Top: SPCE model. Bottom: TIP5P model.
    }
  \label{fig:n10}
\end{figure}

\begin{figure}[!t]
  \centering
  \includegraphics[angle=0,width=7.0cm]{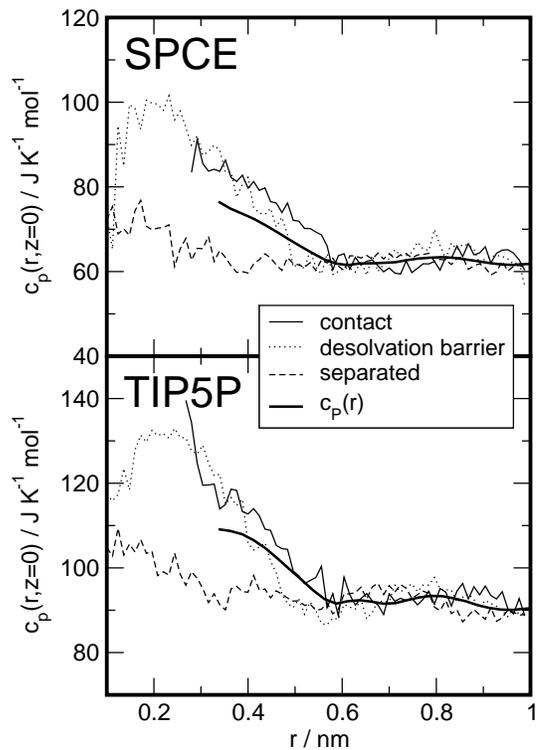}
  \caption{\footnotesize
    Heat capacity of the SPCE water molecules
    around a pair of Xenon particles 
    in the Xe-Xe bisector plane
    $c_P(r,z\!=\!0)$ obtained for a certain Xe-Xe 
    distance interval. The $c_P(r,z\!=\!0)$ data have been obtained as 
    linear fits to the corresponding water potential energies $E_{Water}(r,z=0)$
    using all five temperatures.
    ``contact'': $d_{\rm Xe-Xe}\!\leq\!0.45\,\mbox{nm}$.
    ``desolvation barrier'':
    $0.55\,\mbox{nm}\!\le\!d_{\rm Xe-Xe}\!\leq\!0.65\,\mbox{nm}$.
    ``separated'':
    $0.9\,\mbox{nm}\!\le\!d_{\rm Xe-Xe}\!\leq\!1.0\,\mbox{nm}$.
    For comparison, also the $\Delta c_P(r)+c_{P,\rm bulk}$ data of Figure \ref{fig:n04}
    are given.
    Top: SPCE model. Bottom: TIP5P model.
}
  \label{fig:n11}
\end{figure}

\fi
The observed temperature dependence of the enthalpy profiles is
quantified in terms of a heat capacity profile for the association of two
Xenon particles, shown in Figure \ref{fig:n06}. 
In accordance with the observations of Shimizu and Chan
\cite{Shimizu:2000,Shimizu:2001,Shimizu:2002}, and Rick
\cite{Rick:2003}, we observe a maximum of the heat capacity located at the
desolvation barrier. For the contact state, however, the two water models
employed in our study show a quite different behavior. For the SPCE model
we find almost no difference in the heat capacity between the
contact state and the completely separated state, corresponding to the
observation of Shimizu and Chan \cite{Shimizu:2001,Shimizu:2002} for Methane in TIP4P water.
For the TIP5P model, however, we observe a considerable
 negative heat capacity
contribution at the contact state.

At a first glance this behavior can be rationalized as 
a simple consequence of the overlapping of the hydration
shells in case two hydrophobic particles approach each other.
The positive association enthalpy and entropy corresponds  to a picture of
the  release of shell water molecules
with a lower potential energy and higher ordering 
into the bulk phase. 
However, when taking the temperature dependence of the
$h(r)$ and $s(r)$ profiles into account, two observations are {\em not}
consistent with this simple model considerations:
First, since the hydration shell water molecules have
been shown to exhibit an enhanced heat capacity, a reduced
number of water molecules in the joint hydration shell of a pair
of Xenon atoms in the contact state should also 
lead to a net negative 
heat capacity. This, however, is apparently not the case for
the SPCE model.
The second obvious inconsistency is, of course, the enlarged positive heat capacity
observed at desolvation barrier, found for both water models here,
and even more water models, as discussed in Ref. \cite{Paschek:2004}.

In order to elucidate the observed heat capacity
profile for the Xenon-Xenon association, we calculate
two-dimensional distribution functions describing the 
water density as well as the water heat capacity
around a pair of Xenon particles, found
at a certain distance $d_{\rm Xe-Xe}$. These distributions are shown in
Figures \ref{fig:n08} and \ref{fig:n09}. Figure \ref{fig:n07} illustrates
the definition of distribution functions. We would like to point out
that a conceptually similar plot of 
the molecule densities around a pair of hydrophobic Methane particles
has been recently reported by Gosh et al. \cite{Ghosh:2002,Ghosh:2003}.
If a pair of Xenon atoms is 
found to belong to a certain distance $d_{\rm Xe-Xe}$ interval corresponding
to one of the states ``contact'', ``desolvation barrier'', or ``separated''
(see the figure captions of Figures \ref{fig:n08} and \ref{fig:n09} 
for the corresponding definition intervals), the properties
of the surrounding water molecules (normalized density, potential energy) are mapped 
with respect to their cylinder coordinates $r$ and $z$. $z=0$ is chosen in
such a way that it indicates the bisector plane between the 
two adjacent Xenon particles.

Figure \ref{fig:n08} shows the  normalized cylindrical distribution
functions of the water molecule density  around a pair 
of Xenon atoms $g_{\rm Xe-Xe-W}(r,z)$,
obtained from the SPCE-Xenon simulations at $300\,\mbox{K}$.
In order to provide a more quantitative comparison of the TIP5P and the SPCE models,
we show in Figure \ref{fig:n10} the distribution functions obtained for the
bisector plane $g_{\rm Xe-Xe-W}(r,z=0)$. 
Figure \ref{fig:n08} reveals an enhanced  water molecule density in the range
where the hydration shells of the two Xenon particles overlap.
With $d_{\rm Xe-Xe}\rightarrow0$ the $g_{\rm Xe-Xe-W}(r,z=0)$ function
approaches the $g_{\rm Xe-W}(r)$ pair correlation function. With increasing $d_{\rm Xe-Xe}$ the
first peak starts to increase showing roughly a doubled 
height (with respect to the first peak of $g_{\rm Xe-W}(r)$) of the first
maximum at the desolvation barrier. A further increase of $d_{\rm Xe-Xe}$ leads
again to a decrease in height of the first peak. This is qualitatively
in accord with the observations of Ghosh et al. \cite{Ghosh:2003} obtained
for Methane in TIP3P water. We find that both 
water model models show
qualitatively the same behavior. The TIP5P model, however, exhibits a slightly
stronger increase of the first peak at the desolvation barrier.

The heat capacity of the SPCE water around a pair of Xenon particles $c_P(r,z)$ is shown in 
Figure \ref{fig:n09}. The data were obtained by sampling
the water potential energies as a function of $r$ and $z$ for all
temperatures and subsequent linear fitting of the average
values with respect to the temperature.
The potential energies of the water molecules were
obtained the same way as discussed in the previous sections.
The color coding in Figure \ref{fig:n09} is chosen in such a way that the lower limit
coincides with the average bulk value for waters
configurational heat capacity of
$62.1\,\mbox{J}\,\mbox{K}^{-1}\,\mbox{mol}^{-1}$. 
Since we consider here a system consisting of 8 Xenon particles and 500 water molecules, the bulk
value for $c_{P}$ of water
is slightly larger than the value obtained for pure water discussed in section
\ref{sec:hydration}. The color
spectrum represents a total range of
$30\,\mbox{J}\,\mbox{K}^{-1}\,\mbox{mol}^{-1}$. Figure
\ref{fig:n03} indicates that virtually no
water molecules are found at distances smaller than 
about $0.3\,\mbox{nm}$ to any Xenon site, hence no water $c_P$ data is
available. Therefore this region is represented in Figure \ref{fig:n09}
by the color indicating the lower limit for $c_p$.
In order to provide a more quantitative representation,
Figure \ref{fig:n11} shows the heat capacity of 
the SPCE and TIP5P water molecules in the
bisector plane. 

Figure \ref{fig:n09} reveals the origin the heat capacity maximum 
at the desolvation barrier, shown in Figure \ref{fig:n06}. The water molecules 
adsorbed close to the bisector plane
between the two hydrophobic particles exhibit an
about $60\,\%$ increased heat capacity (in the ``desolvation barrier'' state), 
which apparently overcompensates the
effect of the shrinking of the total hydration shell.
In addition, Figure 
\ref{fig:n09} shows that the increased heat capacity of the hydration shell water
molecules located in the bisector plane persists in the contact state,
still partially compensating the reduced solvent accessible surface. The
differences observed for the contact state of the TIP5P and SPCE models seem 
to be related to the individual strengths of these compensating effects. Figure
\ref{fig:n11} indicates that the SPCE and TIP5P models behave qualitatively
similar. Even the relative changes are of the same size: The 
maximum at the desolvation barrier shows an increase of the heat capacity
of about $40\,\mbox{J}\,\mbox{K}^{-1}\,\mbox{mol}^{-1}$ for 
both water models. However, the TIP5P
water exhibits an about $30\,\mbox{J}\,\mbox{K}^{-1}\,\mbox{mol}^{-1}$
larger bulk heat capacity.
As a consequence, the effect due to the
decrease of the solvent accessible surface has to be stronger 
in case of TIP5P, which might explain the negative
heat capacity at the contact state.
In addition,
the heat capacity effect might as well be
influenced by the second hydration shell of Xenon 
(see Figure \ref{fig:n04}). The heat capacity of the
water in the second hydration shell is clearly 
more strongly affected in case of the TIP5P
model. The contribution of this extended hydration shell should also lead
to a more negative net excess heat capacity.
The observed water heat capacities provide an explanation
for the maximum of the heat capacity profile of two associating
hydrophobic particles. Hence the differences between
the SPCE and TIP5P models seem to be 
related to the delicate balance of two compensating effects: 
the reduction of the solvent accessible surface and 
the increased heat capacity of the
water molecules in the concave part of the joint Xenon-Xenon hydration shell.

\subsection{Hydrogen Bonding}

\ifpretty
\begin{figure}[!b]
  \centering
  \includegraphics[angle=0,width=7.0cm]{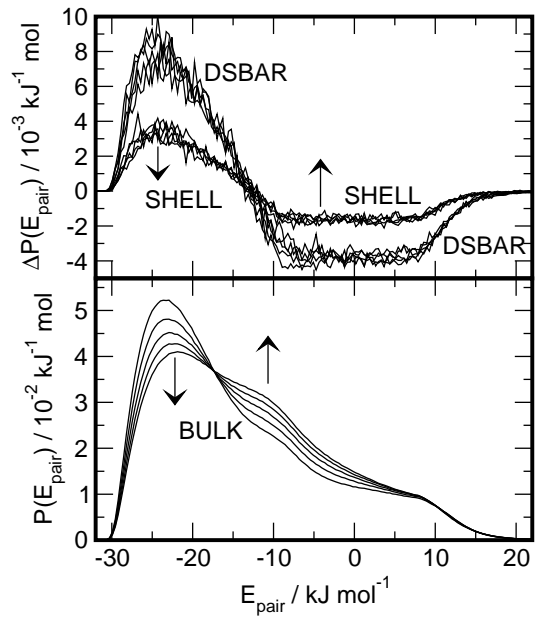}
  \caption{\footnotesize
    Bottom: Probability density of observing a  water-water pair energy
    of water molecules which are separated
    less than $0.35\,\mbox{nm}$, and which are located
    in the ``BULK'' phase.
    Top: Difference between the corresponding probabiliy densities
    of water molecules belonging to the Xenon hydration shell
    (``SHELL''), as well as close to the bisector plane between two Xenon
    particles located at the desolvation barrier 
    (``DSBAR''), and
    the probability density obtained for the bulk (exact definitions 
    are given in the text). The arrows indicate the sequence of curves
    pointing from low to high temperatures.
}
  \label{fig:n12}
\end{figure}

\begin{figure}[!t]
  \centering
  \includegraphics[angle=0,width=7.0cm]{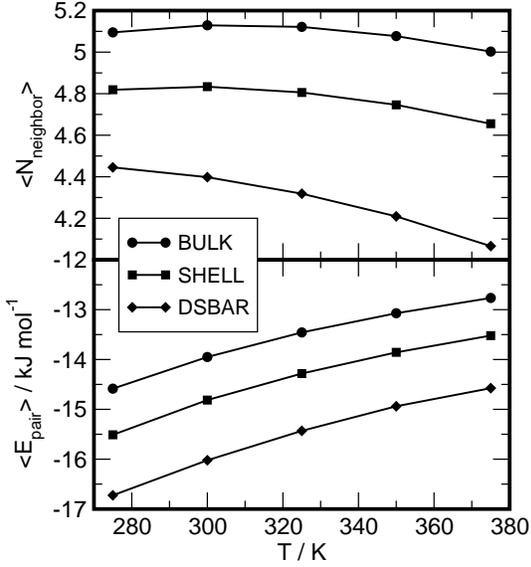}
  \caption{\footnotesize
    Top: 
    Average number of nerarest (water) neighbors of of a SPCE 
    molecule.
    Bottom:
    Average nearest neighbor pair interaction energies of a SPCE water
    molecule belonging to  the states   ``BULK'', ``SHELL'', and ``DSBAR''
    (Definitions are given in the text).
    }
  \label{fig:n13}
\end{figure}

\begin{figure}[!t]
  \centering
  \includegraphics[angle=0,width=7.0cm]{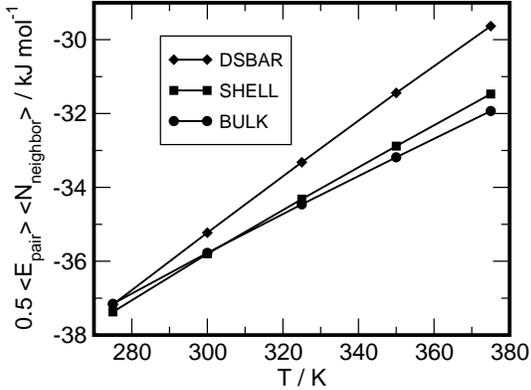}
  \caption{\footnotesize
    Potential energy of a water molecule according to nearest neighbor
    interactions (applying  an oxygen-oxygen cutoff of 
    $r_c\!=\!0.35\,\mbox{nm}$) 
    belonging to the states
    ``BULK'', ``SHELL'', and ``DSBAR''.
    }
  \label{fig:n14}
\end{figure}

\begin{figure}[!t]
  \centering
  \includegraphics[angle=0,width=7.0cm]{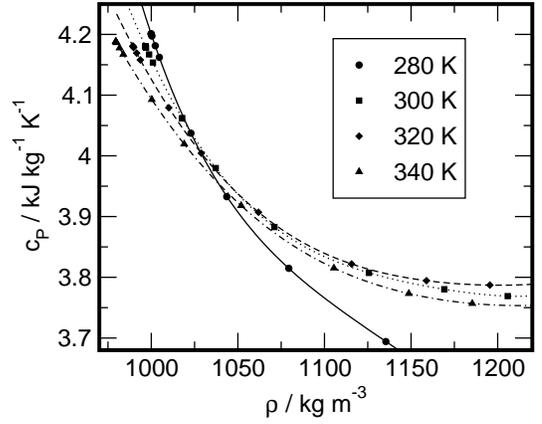}
  \caption{\footnotesize
    Symbols:
    Heat capacity $c_P$ dat of pure water as a function of density
    according  to the IAPWS-95 formulation \cite{Wagner:2002}
    (Data are taken from Ref. \cite{Wagner:2002}).
    Lines: 5th order polynomic fits of the $c_P(\rho)$ data.
    }
  \label{fig:n15}
\end{figure}

\fi
Finally we would like to discuss the
origin of the observed heat capacity
effects. Therefore we calculate the (binding) molecule pair energies 
$E_{\rm pair}$ between 
water molecules which are potentially hydrogen bonded in the sense
that their O-O distance is smaller than $0.35\,\mbox{nm}$ \cite{Geiger:79,Luzar:96:1}. For this
purpose we distinguish between three distinct states: ``BULK'':
The separation from the hydrophobic particle is larger than
$1.0\,\mbox{nm}$.``SHELL'': The water molecules are found to lie within 
a sphere with radius $0.55\,\mbox{nm}$ around the hydrophobic particle.
``DSBAR'': The water molecules are found in the region close to the bisector plane
between two adjacent Xenon particles found within the distance interval 
between $0.55\,\mbox{nm}\!\le\!d_{\rm Xe-Xe}\!\leq\!0.65\,\mbox{nm}$,
in the region defined by  
$r\leq 0.4\,\mbox{nm}$ and $-0.15\,\mbox{nm}\leq z\leq 0.15\,\mbox{nm}$
(see Figures \ref{fig:n07} and \ref{fig:n08} for further explanations). 
For the first two categories we employ 
simulations containing a single Xenon atom, whereas for the ``DSBAR'' category we use the
simulations with 8 Xenon particles. We would like to point out that
the simulations containing 8 Xenon
atoms provide qualitatively similar results for the first two categories.
All data reported in this
section were obtained using the SPCE model, but the TIP5P model is found
to furnish a qualitatively similar picture.

In Figure \ref{fig:n12} the probability densities for obtaining a certain
water-water pair energy $P(E_{\rm pair})$ are given. Comparing the ``SHELL'' state with the
``BULK'' state we find an increased population of states with binding energies
less then approximately $-12\,\mbox{kJ}\,\mbox{mol}^{-1}$. 
Interestingly, the water molecules
in the ``DSBAR'' state show an {\em even more} enhanced population of strongly bound
water molecules. A noteworthy observation is that the temperature variation
of the probability density distributions of the ``SHELL'' and ``DSBAR'' states
with respect to the bulk is found to be only small. At least, significantly
smaller than bringing a water molecule in either the ``SHELL'' or ``DSBAR''
state. Figure \ref{fig:n13} contains the obtained average pair energies,
indicating
that the binding energies according to the ``SHELL'' and ``DSBAR'' states
are shifted to lower values. A consequence of the small temperature variation
of the water-water probability density is that the energy averages 
$\left<E_{\rm pair}\right>$ for the
three different states are running mostly parallel to the bulk data. 
A second parameter controlling
the potential energy of the water molecules is,
of course, the number of possible binding partners, or number of
nearest neighbors $\left< N_{\rm neighbor}\right>$. Figure \ref{fig:n13} therefore
shows the number of nearest neighbors as a function
of temperature.
To obtain a first order approximation to the water potential energy we may just consider the
first solvation shell of each
water molecule, according to
$E\approx1/2 \left< E_{\rm pair}\right> \times \left< N_{\rm
    neighbor}\right>$. 
The corresponding energy data are shown in Figure \ref{fig:n14}.
The temperature dependence reveals that the water
heat capacity effects discussed in
sections \ref{sec:hydration} and \ref{sec:interaction} 
have to be largely attributed to the waters nearest environment.
The heat capacities according 
to the data shown in Figure \ref{fig:n14} are
$52.1\,\mbox{J}\,\mbox{K}^{-1}\,\mbox{mol}^{-1}$, 
$58.9\,\mbox{J}\,\mbox{K}^{-1}\,\mbox{mol}^{-1}$, and
$75.5\,\mbox{J}\,\mbox{K}^{-1}\,\mbox{mol}^{-1}$ 
for ``BULK'', ``SHELL'', and ``DSBAR'', respectively, hence accounting
for about $85\,\%$ of the heat capacity considering the
full environment. We would like to stress two observations:
The reduction of the number of water neighbors, i.e. reduction of the local
water density, leads apparently to a
strengthening of the water/water pair interactions. We would like to
emphasize that this behavior is analogous to
the behavior observed for stretched water 
\cite{Geiger:86,Sciortino:91,Sciortino:92,Poole:92,Poole:93:1}, 
as well as to the behavior
found in low density patches due
to density fluctuations of water at ambient conditions 
\cite{Sciortino:91,Sciortino:92}. 
In Ref. \cite{Sciortino:91} and \cite{Sciortino:92} Sciortino at al. 
showed that a decrease in density is accompanied by a strengthened
water-water binding energy. As mechanism they identify a decreasing
amount of ``fifth neighbor'' configurations,
contributing energetically unfavorable  ``bifurcated'' hydrogen bonds.
The same behavior is observed in the hydrophobic hydration shell.
Hence the water in the ``SHELL'' and ``DSBAR''
states might be considered as {\em locally stretched} water, 
while representing different degrees of stretching.
The increased heat capacity 
with stretching is also consistent with the
experimentally obtained heat capacity of pure water 
based on the IAPWS-95 formulation
according to Wagner and Pru{\ss} 
\cite{Wagner:2002}.The density dependence of the
heat capacities  are shown in Figure \ref{fig:n15} for several temperatures.
As a rough estimate we obtain
$(\partial c_P(\rho)/\partial\rho)_T\!=\!-5.5
\times 10^{3}\,\mbox{kJ}\,\mbox{m}^3\,\mbox{kg}^{-2}\,\mbox{K}^{-1}$
for $T\!=\!300\,\mbox{K}$ around $\rho\!=\!1000\,\mbox{kg}\,\mbox{m}^{-3}$.
The configurational contribution to the heat capacity might 
by approximated as
$c_P(liq.)-c_P(gas)\approx2.27\,\mbox{kJ}\,\mbox{kg}^{-1}\,\mbox{K}^{-1}$.
Hence a density decrease of about $14\%$, as found for the ``DSBAR'' state,
 should lead to an increase
of the configurational heat capacity of about $34\%$. 
A density decrease of about $7\%$, as observed for the ``SHELL'' region,
should  result in a heat capacity 
increase of about $17\%$. Both are quite close to the observed simulation
data. In accordance with recent ideas of Ashbaugh et al.
\cite{Ashbaugh:2002},
the effects of the hydrophobic hydration and interaction discussed here
might therefore simply reflect features of waters unique equation of state.

\ifpretty
\begin{figure}[!t]
  \centering
  \includegraphics[angle=0,width=7.0cm]{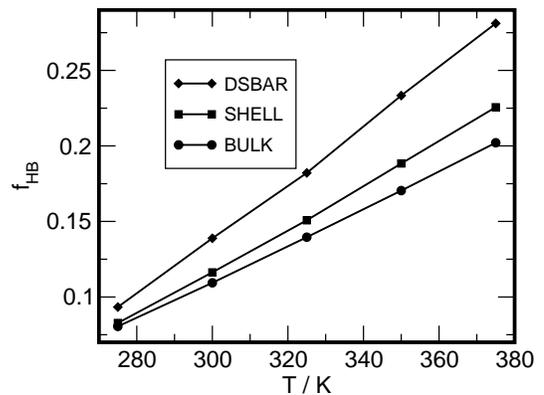}
  \caption{\footnotesize
    Fraction of broken hydrogen bonds $f_{HB}$ for water molecules
    belonging to the states
    ``BULK'', ``SHELL'', and ``DSBAR''.
    }
  \label{fig:n16}
\end{figure}

\fi
Our simulations indicate that the 
water-water binding energy-decrease counterbalances
(and at lower temperature even overcompensates) the effect of the 
dilution of water molecules in the vicinity of hydrophobic particles. 
With increasing temperature the apparent enhanced disintegration of the 
water network (number of nearest neighbors in Figure \ref{fig:n14}) close
to hydrophobic particles is therefore responsible for the observed heat capacity
effects. 
To illustrate this, we finally discuss hydrogen bonding based on a topological
criterion. A pair of water molecules is considered to by ``hydrogen bonded'' when
the O-O distance is smaller than $0.35\,\mbox{nm}$ and the 
$\mbox{H}-\mbox{O}\cdots\mbox{O}$ 
angle is smaller than $30^\circ$ \cite{Starr:99,Luzar:96:1}.
 Each water hydrogen is considered to be
a hydrogen bond donor and $f_{HB}$ denotes the fraction of hydrogen bond
donors which are {\em not} involved in a hydrogen bond, or the
fraction of broken hydrogen bonds. $f_{HB}$ as a function of temperature is
shown in Figure \ref{fig:n16} for water molecules belonging to 
different states. 
In parallel with observed increasing potential energies, the
increasing temperature leads to an enhanced breaking of hydrogen bonds. This
effect is found to be enhanced in the hydration shell of the hydrophobic
particles and particularly strong for water molecules in the ``DSBAR'' state.


\section{CONCLUSIONS}

We have shown that for simple hydrophobic solutes the 
observed positive excess heat capacity of solvation is 
largely (about 90\%) determined by an increase of
the heat capacity of the water in the first hydration shell. 
The effect is mainly attributed to
the altered potential energy of the water molecules in the
first hydration shell, changing their energy state 
from ``attractive'' at lower temperatures to ``repulsive'' at higher
temperatures. 
The  two model systems studied, SPCE and TIP5P water, show a qualitatively
similar behavior, although the effect is found to be quantitatively stronger
in case of the TIP5P model.

Our model calculations, in accordance with the calculations of Shimizu and Chan
\cite{Shimizu:2000,Shimizu:2001,Shimizu:2002}, as well as Rick
\cite{Rick:2003} and Paschek \cite{Paschek:2004},
indicate that 
the heat capacity for the association of two hydrophobic particles
exhibits a maximum located at the desolvation barrier.
Our simulations indicate that this seems to 
be a  characteristic feature of the hydrophobic
interaction of small apolar particles. The  observed behavior
is found to be a counterplay of two effects: The reduced solvent accessible
surface, and the increased heat capacity of water molecules located in the
gap between the two hydrophobic particles. In case of the 
SPCE model, the water molecules located in this bridging situation show a
heat capacity increase of about $60\,\%$ compared to the bulk, apparently
overcompensating the effect of a reduced solvent accessible surface.
The differences found for the SPCE and TIP5P models in the contact state
are accordingly related to quantitative differences in strength of this two
compensating effects.

A detailed analysis of the water-water pair interactions in the different
states (bulk, hydration shell, joint hydration shell of two Xenon atoms)
reveals that the observed heat capacity effects can be rationalized
as a counterbalance of strengthened hydrogen bonds and an enhanced
disintegration of the hydrogen bond network with increasing temperature.
The reduced number of water neighbors in the different parts of the 
joint hydrophobic hydration shell might be interpreted as a {\em locally
stretched} water/hydrogen bond network. A comparison with experimental data 
indicates that the observed heat capacity effects
have the same order of magnitude as it would be  expected for pure water with
an approximately equally reduced density.


\section*{ACKNOWLEGEMENT}

I would like to thank Alfons Geiger and Ivan Brovchenko for insightful
discussions. The analogy between hydration shell water and
stretched water has been originally suggested by Alfons Geiger. 
Financing by the Deutsche Forschungsgemeinschaft
(DFG Forschergruppe 436) is gratefully acknowledged. 


\begin{thebibliography}{68}
\expandafter\ifx\csname natexlab\endcsname\relax\def\natexlab#1{#1}\fi
\expandafter\ifx\csname bibnamefont\endcsname\relax
  \def\bibnamefont#1{#1}\fi
\expandafter\ifx\csname bibfnamefont\endcsname\relax
  \def\bibfnamefont#1{#1}\fi
\expandafter\ifx\csname citenamefont\endcsname\relax
  \def\citenamefont#1{#1}\fi
\expandafter\ifx\csname url\endcsname\relax
  \def\url#1{\texttt{#1}}\fi
\expandafter\ifx\csname urlprefix\endcsname\relax\def\urlprefix{URL }\fi
\providecommand{\bibinfo}[2]{#2}
\providecommand{\eprint}[2][]{\url{#2}}

\bibitem[{\citenamefont{Tanford}(1980)}]{Tanford}
\bibinfo{author}{\bibfnamefont{C.}~\bibnamefont{Tanford}},
  \emph{\bibinfo{title}{The Hydrophobic Effect: Formation of Micelles and
  Biological Membranes}} (\bibinfo{publisher}{John Wiley \& Sons},
  \bibinfo{address}{New York}, \bibinfo{year}{1980}), \bibinfo{edition}{2nd}
  ed.

\bibitem[{\citenamefont{Ben-Naim}(1980)}]{Ben-Naim:Hydrophobic}
\bibinfo{author}{\bibfnamefont{A.}~\bibnamefont{Ben-Naim}},
  \emph{\bibinfo{title}{Hydrophobic Interactions}} (\bibinfo{publisher}{Plenum
  Press}, \bibinfo{address}{New York}, \bibinfo{year}{1980}).

\bibitem[{\citenamefont{Southall et~al.}(2002)\citenamefont{Southall, Dill, and
  Haymet}}]{Southall:2002}
\bibinfo{author}{\bibfnamefont{N.~T.} \bibnamefont{Southall}},
  \bibinfo{author}{\bibfnamefont{K.~A.} \bibnamefont{Dill}}, \bibnamefont{and}
  \bibinfo{author}{\bibfnamefont{A.~D.~J.} \bibnamefont{Haymet}},
  \bibinfo{journal}{J. Phys. Chem. B} \textbf{\bibinfo{volume}{106}},
  \bibinfo{pages}{521} (\bibinfo{year}{2002}).

\bibitem[{\citenamefont{Muller}(1990)}]{Muller:90}
\bibinfo{author}{\bibfnamefont{N.}~\bibnamefont{Muller}},
  \bibinfo{journal}{Acc. Chem. Res.} \textbf{\bibinfo{volume}{23}},
  \bibinfo{pages}{23} (\bibinfo{year}{1990}).

\bibitem[{\citenamefont{Lee and Graziano}(1995)}]{LeeB:95}
\bibinfo{author}{\bibfnamefont{B.}~\bibnamefont{Lee}} \bibnamefont{and}
  \bibinfo{author}{\bibfnamefont{G.}~\bibnamefont{Graziano}},
  \bibinfo{journal}{J. Am. Chem. Soc.} \textbf{\bibinfo{volume}{118}},
  \bibinfo{pages}{5165} (\bibinfo{year}{1995}).

\bibitem[{\citenamefont{Moelbert and {De Los
  Rios}}(2003{\natexlab{a}})}]{Moelbert:2003:2}
\bibinfo{author}{\bibfnamefont{S.}~\bibnamefont{Moelbert}} \bibnamefont{and}
  \bibinfo{author}{\bibfnamefont{P.}~\bibnamefont{{De Los Rios}}},
  \bibinfo{journal}{J. Chem. Phys.} \textbf{\bibinfo{volume}{119}},
  \bibinfo{pages}{7988} (\bibinfo{year}{2003}{\natexlab{a}}).

\bibitem[{\citenamefont{Moelbert and {De Los
  Rios}}(2003{\natexlab{b}})}]{Moelbert:2003:1}
\bibinfo{author}{\bibfnamefont{S.}~\bibnamefont{Moelbert}} \bibnamefont{and}
  \bibinfo{author}{\bibfnamefont{P.}~\bibnamefont{{De Los Rios}}},
  \bibinfo{journal}{Macromolecules} \textbf{\bibinfo{volume}{36}},
  \bibinfo{pages}{5845} (\bibinfo{year}{2003}{\natexlab{b}}).

\bibitem[{\citenamefont{Smith et~al.}(1992)\citenamefont{Smith, Zhang, and
  Haymet}}]{Smith.D:92}
\bibinfo{author}{\bibfnamefont{D.~E.} \bibnamefont{Smith}},
  \bibinfo{author}{\bibfnamefont{L.}~\bibnamefont{Zhang}}, \bibnamefont{and}
  \bibinfo{author}{\bibfnamefont{A.~D.~J.} \bibnamefont{Haymet}},
  \bibinfo{journal}{J. Am. Chem. Soc.} \textbf{\bibinfo{volume}{114}},
  \bibinfo{pages}{5875} (\bibinfo{year}{1992}).

\bibitem[{\citenamefont{Smith and Haymet}(1993)}]{Smith.D:93}
\bibinfo{author}{\bibfnamefont{D.~E.} \bibnamefont{Smith}} \bibnamefont{and}
  \bibinfo{author}{\bibfnamefont{A.~D.~J.} \bibnamefont{Haymet}},
  \bibinfo{journal}{J. Chem. Phys.} \textbf{\bibinfo{volume}{98}},
  \bibinfo{pages}{6445} (\bibinfo{year}{1993}).

\bibitem[{\citenamefont{Geiger et~al.}(1979)\citenamefont{Geiger, Rahman, and
  Stillinger}}]{Geiger:79}
\bibinfo{author}{\bibfnamefont{A.}~\bibnamefont{Geiger}},
  \bibinfo{author}{\bibfnamefont{A.}~\bibnamefont{Rahman}}, \bibnamefont{and}
  \bibinfo{author}{\bibfnamefont{F.~H.} \bibnamefont{Stillinger}},
  \bibinfo{journal}{J. Chem. Phys} \textbf{\bibinfo{volume}{70}},
  \bibinfo{pages}{263} (\bibinfo{year}{1979}).

\bibitem[{\citenamefont{Zichi and Rossky}(1985)}]{Zichi:85}
\bibinfo{author}{\bibfnamefont{D.~A.} \bibnamefont{Zichi}} \bibnamefont{and}
  \bibinfo{author}{\bibfnamefont{P.~J.} \bibnamefont{Rossky}},
  \bibinfo{journal}{J. Chem. Phys.} \textbf{\bibinfo{volume}{83}},
  \bibinfo{pages}{797} (\bibinfo{year}{1985}).

\bibitem[{\citenamefont{Pearlman}(1993)}]{Pearlman:93}
\bibinfo{author}{\bibfnamefont{D.~A.} \bibnamefont{Pearlman}},
  \bibinfo{journal}{J. Chem. Phys.} \textbf{\bibinfo{volume}{98}},
  \bibinfo{pages}{8946} (\bibinfo{year}{1993}).

\bibitem[{\citenamefont{{van Belle} and Wodak}(1993)}]{Belle:93}
\bibinfo{author}{\bibfnamefont{D.}~\bibnamefont{{van Belle}}} \bibnamefont{and}
  \bibinfo{author}{\bibfnamefont{S.~J.} \bibnamefont{Wodak}},
  \bibinfo{journal}{J. Am. Chem. Soc} \textbf{\bibinfo{volume}{115}},
  \bibinfo{pages}{647} (\bibinfo{year}{1993}).

\bibitem[{\citenamefont{Dang}(1994)}]{Dang:94}
\bibinfo{author}{\bibfnamefont{L.~X.} \bibnamefont{Dang}}, \bibinfo{journal}{J.
  Chem. Phys.} \textbf{\bibinfo{volume}{100}}, \bibinfo{pages}{9032}
  (\bibinfo{year}{1994}).

\bibitem[{\citenamefont{Forsman and J\"onsson}(1994)}]{Forsman:94}
\bibinfo{author}{\bibfnamefont{J.}~\bibnamefont{Forsman}} \bibnamefont{and}
  \bibinfo{author}{\bibfnamefont{B.}~\bibnamefont{J\"onsson}},
  \bibinfo{journal}{J. Chem. Phys.} \textbf{\bibinfo{volume}{101}},
  \bibinfo{pages}{5116} (\bibinfo{year}{1994}).

\bibitem[{\citenamefont{L\"udemann et~al.}(1996)\citenamefont{L\"udemann,
  Schreiber, Abseher, and Steinhauser}}]{Luedemann:96}
\bibinfo{author}{\bibfnamefont{S.}~\bibnamefont{L\"udemann}},
  \bibinfo{author}{\bibfnamefont{H.}~\bibnamefont{Schreiber}},
  \bibinfo{author}{\bibfnamefont{R.}~\bibnamefont{Abseher}}, \bibnamefont{and}
  \bibinfo{author}{\bibfnamefont{O.}~\bibnamefont{Steinhauser}},
  \bibinfo{journal}{J. Chem. Phys.} \textbf{\bibinfo{volume}{104}},
  \bibinfo{pages}{286} (\bibinfo{year}{1996}).

\bibitem[{\citenamefont{Skipper et~al.}(1996)\citenamefont{Skipper, Bridgeman,
  Buckingham, and Mancera}}]{Skipper:96}
\bibinfo{author}{\bibfnamefont{N.~T.} \bibnamefont{Skipper}},
  \bibinfo{author}{\bibfnamefont{C.~H.} \bibnamefont{Bridgeman}},
  \bibinfo{author}{\bibfnamefont{A.~D.} \bibnamefont{Buckingham}},
  \bibnamefont{and} \bibinfo{author}{\bibfnamefont{R.~L.}
  \bibnamefont{Mancera}}, \bibinfo{journal}{Faraday Discuss.}
  \textbf{\bibinfo{volume}{103}}, \bibinfo{pages}{141} (\bibinfo{year}{1996}).

\bibitem[{\citenamefont{Young and {Brooks III.}}(1997)}]{Young:97}
\bibinfo{author}{\bibfnamefont{W.~S.} \bibnamefont{Young}} \bibnamefont{and}
  \bibinfo{author}{\bibfnamefont{C.~L.} \bibnamefont{{Brooks III.}}},
  \bibinfo{journal}{J. Chem. Phys.} \textbf{\bibinfo{volume}{106}},
  \bibinfo{pages}{9265} (\bibinfo{year}{1997}).

\bibitem[{\citenamefont{Hummer et~al.}(1996)\citenamefont{Hummer, Garde,
  Garcia, Pohorille, and Pratt}}]{Hummer:96:1}
\bibinfo{author}{\bibfnamefont{G.}~\bibnamefont{Hummer}},
  \bibinfo{author}{\bibfnamefont{S.}~\bibnamefont{Garde}},
  \bibinfo{author}{\bibfnamefont{A.~E.} \bibnamefont{Garcia}},
  \bibinfo{author}{\bibfnamefont{A.}~\bibnamefont{Pohorille}},
  \bibnamefont{and} \bibinfo{author}{\bibfnamefont{L.~R.} \bibnamefont{Pratt}},
  \bibinfo{journal}{Proc. Natl. Acad. Sci. USA} \textbf{\bibinfo{volume}{93}},
  \bibinfo{pages}{8951} (\bibinfo{year}{1996}).

\bibitem[{\citenamefont{L\"udemann et~al.}(1997)\citenamefont{L\"udemann,
  Abseher, Schreiber, and Steinhauser}}]{Luedemann:97}
\bibinfo{author}{\bibfnamefont{S.}~\bibnamefont{L\"udemann}},
  \bibinfo{author}{\bibfnamefont{R.}~\bibnamefont{Abseher}},
  \bibinfo{author}{\bibfnamefont{H.}~\bibnamefont{Schreiber}},
  \bibnamefont{and}
  \bibinfo{author}{\bibfnamefont{O.}~\bibnamefont{Steinhauser}},
  \bibinfo{journal}{J. Am. Chem. Soc.} \textbf{\bibinfo{volume}{119}},
  \bibinfo{pages}{4206} (\bibinfo{year}{1997}).

\bibitem[{\citenamefont{Rick and Berne}(1997)}]{Rick:97}
\bibinfo{author}{\bibfnamefont{S.~W.} \bibnamefont{Rick}} \bibnamefont{and}
  \bibinfo{author}{\bibfnamefont{B.~J.} \bibnamefont{Berne}},
  \bibinfo{journal}{J. Phys. Chem. B} \textbf{\bibinfo{volume}{101}},
  \bibinfo{pages}{10488} (\bibinfo{year}{1997}).

\bibitem[{\citenamefont{Hummer}(2001)}]{Hummer:2001}
\bibinfo{author}{\bibfnamefont{G.}~\bibnamefont{Hummer}}, \bibinfo{journal}{J.
  Chem. Phys.} \textbf{\bibinfo{volume}{114}}, \bibinfo{pages}{7330}
  (\bibinfo{year}{2001}).

\bibitem[{\citenamefont{Shimizu and Chan}(2000)}]{Shimizu:2000}
\bibinfo{author}{\bibfnamefont{S.}~\bibnamefont{Shimizu}} \bibnamefont{and}
  \bibinfo{author}{\bibfnamefont{H.~S.} \bibnamefont{Chan}},
  \bibinfo{journal}{J. Chem. Phys.} \textbf{\bibinfo{volume}{113}},
  \bibinfo{pages}{4683} (\bibinfo{year}{2000}).

\bibitem[{\citenamefont{Rick}(2000)}]{Rick:2000}
\bibinfo{author}{\bibfnamefont{S.~W.} \bibnamefont{Rick}}, \bibinfo{journal}{J.
  Phys. Chem. B} \textbf{\bibinfo{volume}{104}}, \bibinfo{pages}{6884}
  (\bibinfo{year}{2000}).

\bibitem[{\citenamefont{Shimizu and Chan}(2001)}]{Shimizu:2001}
\bibinfo{author}{\bibfnamefont{S.}~\bibnamefont{Shimizu}} \bibnamefont{and}
  \bibinfo{author}{\bibfnamefont{H.~S.} \bibnamefont{Chan}},
  \bibinfo{journal}{J. Am. Chem. Soc} \textbf{\bibinfo{volume}{123}},
  \bibinfo{pages}{2083} (\bibinfo{year}{2001}).

\bibitem[{\citenamefont{Shimizu and Chan}(2002)}]{Shimizu:2002}
\bibinfo{author}{\bibfnamefont{S.}~\bibnamefont{Shimizu}} \bibnamefont{and}
  \bibinfo{author}{\bibfnamefont{H.}~\bibnamefont{Chan}},
  \bibinfo{journal}{Proteins: Struct., Funct., Genet.} pp.
  \bibinfo{pages}{560--566} (\bibinfo{year}{2002}).

\bibitem[{\citenamefont{Ghosh et~al.}(2001)\citenamefont{Ghosh, Garcia, and
  Garde}}]{Ghosh:2001}
\bibinfo{author}{\bibfnamefont{T.}~\bibnamefont{Ghosh}},
  \bibinfo{author}{\bibfnamefont{A.~E.} \bibnamefont{Garcia}},
  \bibnamefont{and} \bibinfo{author}{\bibfnamefont{S.}~\bibnamefont{Garde}},
  \bibinfo{journal}{J. Am. Chem. Soc.} \textbf{\bibinfo{volume}{123}},
  \bibinfo{pages}{10997} (\bibinfo{year}{2001}).

\bibitem[{\citenamefont{Ghosh et~al.}(2002)\citenamefont{Ghosh, Garcia, and
  Garde}}]{Ghosh:2002}
\bibinfo{author}{\bibfnamefont{T.}~\bibnamefont{Ghosh}},
  \bibinfo{author}{\bibfnamefont{A.~E.} \bibnamefont{Garcia}},
  \bibnamefont{and} \bibinfo{author}{\bibfnamefont{S.}~\bibnamefont{Garde}},
  \bibinfo{journal}{J. Chem. Phys} \textbf{\bibinfo{volume}{116}},
  \bibinfo{pages}{2480} (\bibinfo{year}{2002}).

\bibitem[{\citenamefont{Ghosh et~al.}(2003)\citenamefont{Ghosh, Garcia, and
  Garde}}]{Ghosh:2003}
\bibinfo{author}{\bibfnamefont{T.}~\bibnamefont{Ghosh}},
  \bibinfo{author}{\bibfnamefont{A.~E.} \bibnamefont{Garcia}},
  \bibnamefont{and} \bibinfo{author}{\bibfnamefont{S.}~\bibnamefont{Garde}},
  \bibinfo{journal}{J. Phys. Chem. B} \textbf{\bibinfo{volume}{107}},
  \bibinfo{pages}{612} (\bibinfo{year}{2003}).

\bibitem[{\citenamefont{Rick}(2003)}]{Rick:2003}
\bibinfo{author}{\bibfnamefont{S.~W.} \bibnamefont{Rick}}, \bibinfo{journal}{J.
  Chem. Phys. B} \textbf{\bibinfo{volume}{107}}, \bibinfo{pages}{9853}
  (\bibinfo{year}{2003}).

\bibitem[{\citenamefont{Lee and Richards}(1971)}]{Lee:71}
\bibinfo{author}{\bibfnamefont{B.}~\bibnamefont{Lee}} \bibnamefont{and}
  \bibinfo{author}{\bibfnamefont{F.~M.} \bibnamefont{Richards}},
  \bibinfo{journal}{J. Mol. Biol.} \textbf{\bibinfo{volume}{55}},
  \bibinfo{pages}{379} (\bibinfo{year}{1971}).

\bibitem[{\citenamefont{Makhatadze and Privalov}(1995)}]{Makhatadze:95}
\bibinfo{author}{\bibfnamefont{G.~I.} \bibnamefont{Makhatadze}}
  \bibnamefont{and} \bibinfo{author}{\bibfnamefont{P.~L.}
  \bibnamefont{Privalov}}, \bibinfo{journal}{Adv. Protein Chem.}
  \textbf{\bibinfo{volume}{47}}, \bibinfo{pages}{307} (\bibinfo{year}{1995}).

\bibitem[{\citenamefont{Pratt}(2003)}]{PrattRev:2002}
\bibinfo{author}{\bibfnamefont{L.~R.} \bibnamefont{Pratt}},
  \bibinfo{journal}{Annu. Rev. Phys. Chem.} \textbf{\bibinfo{volume}{53}},
  \bibinfo{pages}{409} (\bibinfo{year}{2003}).

\bibitem[{\citenamefont{Widom et~al.}(2003)\citenamefont{Widom, Bhimalapuram,
  and Koga}}]{Widom:2003}
\bibinfo{author}{\bibfnamefont{B.}~\bibnamefont{Widom}},
  \bibinfo{author}{\bibfnamefont{P.}~\bibnamefont{Bhimalapuram}},
  \bibnamefont{and} \bibinfo{author}{\bibfnamefont{K.}~\bibnamefont{Koga}},
  \bibinfo{journal}{Phys. Chem. Chem. Phys.} \textbf{\bibinfo{volume}{5}},
  \bibinfo{pages}{3085} (\bibinfo{year}{2003}).

\bibitem[{\citenamefont{Smith and Haymet}(2003)}]{Smith:2003}
\bibinfo{author}{\bibfnamefont{D.~E.} \bibnamefont{Smith}} \bibnamefont{and}
  \bibinfo{author}{\bibfnamefont{A.~D.~J.} \bibnamefont{Haymet}}, in
  \emph{\bibinfo{booktitle}{Reviews in Computational Chemistry}}, edited by
  \bibinfo{editor}{\bibfnamefont{K.~B.} \bibnamefont{Lipkowitz}},
  \bibinfo{editor}{\bibfnamefont{R.}~\bibnamefont{Larter}}, \bibnamefont{and}
  \bibinfo{editor}{\bibfnamefont{T.~R.} \bibnamefont{Cundari}}
  (\bibinfo{publisher}{Wiley-VCH}, \bibinfo{address}{New York},
  \bibinfo{year}{2003}), vol.~\bibinfo{volume}{19}, chap.~\bibinfo{chapter}{2},
  pp. \bibinfo{pages}{44--77}.

\bibitem[{\citenamefont{Paschek}(2004)}]{Paschek:2004}
\bibinfo{author}{\bibfnamefont{D.}~\bibnamefont{Paschek}}, \bibinfo{journal}{J.
  Chem. Phys.}  (\bibinfo{year}{2004}), \bibinfo{note}{accepted for publication
  (Preprint: cond-mat/0312252)}.

\bibitem[{\citenamefont{Nos\'e}(1984)}]{Nose:84}
\bibinfo{author}{\bibfnamefont{S.}~\bibnamefont{Nos\'e}},
  \bibinfo{journal}{Mol. Phys.} \textbf{\bibinfo{volume}{52}},
  \bibinfo{pages}{255} (\bibinfo{year}{1984}).

\bibitem[{\citenamefont{Hoover}(1985)}]{Hoover:85}
\bibinfo{author}{\bibfnamefont{W.~G.} \bibnamefont{Hoover}},
  \bibinfo{journal}{Phys. Rev. A} \textbf{\bibinfo{volume}{31}}
  (\bibinfo{year}{1985}).

\bibitem[{\citenamefont{M.~Parrinello}(1981)}]{Parrinello:81}
\bibinfo{author}{\bibfnamefont{A.~R.} \bibnamefont{M.~Parrinello}},
  \bibinfo{journal}{J. Appl. Phys.} \textbf{\bibinfo{volume}{52}},
  \bibinfo{pages}{7182} (\bibinfo{year}{1981}).

\bibitem[{\citenamefont{Nos\'e and Klein}(1983)}]{Nose:83}
\bibinfo{author}{\bibfnamefont{S.}~\bibnamefont{Nos\'e}} \bibnamefont{and}
  \bibinfo{author}{\bibfnamefont{M.~L.} \bibnamefont{Klein}},
  \bibinfo{journal}{Mol. Phys.} \textbf{\bibinfo{volume}{50}},
  \bibinfo{pages}{1055} (\bibinfo{year}{1983}).

\bibitem[{\citenamefont{Essmann et~al.}(1995)\citenamefont{Essmann, Perera,
  Berkowitz, Darden, Lee, and Pedersen}}]{Essmann:95}
\bibinfo{author}{\bibfnamefont{U.}~\bibnamefont{Essmann}},
  \bibinfo{author}{\bibfnamefont{L.}~\bibnamefont{Perera}},
  \bibinfo{author}{\bibfnamefont{M.~L.} \bibnamefont{Berkowitz}},
  \bibinfo{author}{\bibfnamefont{T.~A.} \bibnamefont{Darden}},
  \bibinfo{author}{\bibfnamefont{H.}~\bibnamefont{Lee}}, \bibnamefont{and}
  \bibinfo{author}{\bibfnamefont{L.~G.} \bibnamefont{Pedersen}},
  \bibinfo{journal}{J. Chem. Phys.} \textbf{\bibinfo{volume}{103}},
  \bibinfo{pages}{8577} (\bibinfo{year}{1995}).

\bibitem[{\citenamefont{Miyamoto and Kollman}(1992)}]{Miyamoto:92}
\bibinfo{author}{\bibfnamefont{S.}~\bibnamefont{Miyamoto}} \bibnamefont{and}
  \bibinfo{author}{\bibfnamefont{P.~A.} \bibnamefont{Kollman}},
  \bibinfo{journal}{J. Comp. Chem.} \textbf{\bibinfo{volume}{13}},
  \bibinfo{pages}{952} (\bibinfo{year}{1992}).

\bibitem[{\citenamefont{Lindahl et~al.}(2001)\citenamefont{Lindahl, Hess, and
  {van der Spoel}}}]{gmxpaper}
\bibinfo{author}{\bibfnamefont{E.}~\bibnamefont{Lindahl}},
  \bibinfo{author}{\bibfnamefont{B.}~\bibnamefont{Hess}}, \bibnamefont{and}
  \bibinfo{author}{\bibfnamefont{D.}~\bibnamefont{{van der Spoel}}},
  \bibinfo{journal}{J. Mol. Mod.} \textbf{\bibinfo{volume}{7}},
  \bibinfo{pages}{306} (\bibinfo{year}{2001}).

\bibitem[{\citenamefont{van~der Spoel et~al.}(2001)\citenamefont{van~der Spoel,
  van Buuren, Apol, Meulen\-hoff, Tieleman, Sij\-bers, Hess, Feenstra, Lindahl,
  van Drunen et~al.}}]{gmx31}
\bibinfo{author}{\bibfnamefont{D.}~\bibnamefont{van~der Spoel}},
  \bibinfo{author}{\bibfnamefont{A.~R.} \bibnamefont{van Buuren}},
  \bibinfo{author}{\bibfnamefont{E.}~\bibnamefont{Apol}},
  \bibinfo{author}{\bibfnamefont{P.~J.} \bibnamefont{Meulen\-hoff}},
  \bibinfo{author}{\bibfnamefont{D.~P.} \bibnamefont{Tieleman}},
  \bibinfo{author}{\bibfnamefont{A.~L. T.~M.} \bibnamefont{Sij\-bers}},
  \bibinfo{author}{\bibfnamefont{B.}~\bibnamefont{Hess}},
  \bibinfo{author}{\bibfnamefont{K.~A.} \bibnamefont{Feenstra}},
  \bibinfo{author}{\bibfnamefont{E.}~\bibnamefont{Lindahl}},
  \bibinfo{author}{\bibfnamefont{R.}~\bibnamefont{van Drunen}},
  \bibnamefont{et~al.}, \emph{\bibinfo{title}{Gromacs {U}ser {M}anual version
  3.1}}, \bibinfo{address}{Nij\-enborgh 4, 9747 AG Groningen, The Netherlands.
  Internet: http://www.gromacs.org} (\bibinfo{year}{2001}).

\bibitem[{\citenamefont{Flyvbjerg and Petersen}(1989)}]{Flyvbjerg:89}
\bibinfo{author}{\bibfnamefont{H.}~\bibnamefont{Flyvbjerg}} \bibnamefont{and}
  \bibinfo{author}{\bibfnamefont{H.~G.} \bibnamefont{Petersen}},
  \bibinfo{journal}{J. Chem. Phys.} \textbf{\bibinfo{volume}{91}},
  \bibinfo{pages}{461} (\bibinfo{year}{1989}).

\bibitem[{\citenamefont{Berendsen et~al.}(1984)\citenamefont{Berendsen, Postma,
  van Gunsteren, DiNola, and Haak}}]{Berendsen:84}
\bibinfo{author}{\bibfnamefont{H.~J.~C.} \bibnamefont{Berendsen}},
  \bibinfo{author}{\bibfnamefont{J.~P.~M.} \bibnamefont{Postma}},
  \bibinfo{author}{\bibfnamefont{W.~F.} \bibnamefont{van Gunsteren}},
  \bibinfo{author}{\bibfnamefont{A.}~\bibnamefont{DiNola}}, \bibnamefont{and}
  \bibinfo{author}{\bibfnamefont{J.~R.} \bibnamefont{Haak}},
  \bibinfo{journal}{J. Chem. Phys.} \textbf{\bibinfo{volume}{81}},
  \bibinfo{pages}{3684} (\bibinfo{year}{1984}).

\bibitem[{\citenamefont{Berendsen et~al.}(1987)\citenamefont{Berendsen,
  Grigera, and Straatsma}}]{Berendsen:87}
\bibinfo{author}{\bibfnamefont{H.~J.~C.} \bibnamefont{Berendsen}},
  \bibinfo{author}{\bibfnamefont{J.~R.} \bibnamefont{Grigera}},
  \bibnamefont{and} \bibinfo{author}{\bibfnamefont{T.~P.}
  \bibnamefont{Straatsma}}, \bibinfo{journal}{J. Phys. Chem.}
  \textbf{\bibinfo{volume}{91}}, \bibinfo{pages}{6269} (\bibinfo{year}{1987}).

\bibitem[{\citenamefont{Guillot and Guissani}(1993)}]{Guillot:93}
\bibinfo{author}{\bibfnamefont{B.}~\bibnamefont{Guillot}} \bibnamefont{and}
  \bibinfo{author}{\bibfnamefont{Y.}~\bibnamefont{Guissani}},
  \bibinfo{journal}{J. Chem. Phys.} \textbf{\bibinfo{volume}{99}},
  \bibinfo{pages}{8075} (\bibinfo{year}{1993}).

\bibitem[{\citenamefont{Mahoney and Jorgensen}(2000)}]{Mahoney:2000}
\bibinfo{author}{\bibfnamefont{M.~W.} \bibnamefont{Mahoney}} \bibnamefont{and}
  \bibinfo{author}{\bibfnamefont{W.~L.} \bibnamefont{Jorgensen}},
  \bibinfo{journal}{J. Chem. Phys.} \textbf{\bibinfo{volume}{112}},
  \bibinfo{pages}{8910} (\bibinfo{year}{2000}).

\bibitem[{\citenamefont{Widom}(1963)}]{Widom:63}
\bibinfo{author}{\bibfnamefont{B.}~\bibnamefont{Widom}}, \bibinfo{journal}{J.
  Chem. Phys.} \textbf{\bibinfo{volume}{39}}, \bibinfo{pages}{2808}
  (\bibinfo{year}{1963}).

\bibitem[{\citenamefont{Frenkel and Smit}(2002)}]{FrenkelSmit}
\bibinfo{author}{\bibfnamefont{D.}~\bibnamefont{Frenkel}} \bibnamefont{and}
  \bibinfo{author}{\bibfnamefont{B.}~\bibnamefont{Smit}},
  \emph{\bibinfo{title}{Understanding Molecular Simulation --- From Algorithms
  to Applications}} (\bibinfo{publisher}{Academic Press}, \bibinfo{address}{San
  Diego}, \bibinfo{year}{2002}), \bibinfo{edition}{2nd} ed.

\bibitem[{\citenamefont{Deitrick et~al.}(1989)\citenamefont{Deitrick, Scriven,
  and Davis}}]{Deitrick:89}
\bibinfo{author}{\bibfnamefont{G.~L.} \bibnamefont{Deitrick}},
  \bibinfo{author}{\bibfnamefont{L.~E.} \bibnamefont{Scriven}},
  \bibnamefont{and} \bibinfo{author}{\bibfnamefont{H.~T.} \bibnamefont{Davis}},
  \bibinfo{journal}{J. Chem. Phys.} \textbf{\bibinfo{volume}{90}},
  \bibinfo{pages}{2370} (\bibinfo{year}{1989}).

\bibitem[{\citenamefont{Deitrick et~al.}(1992)\citenamefont{Deitrick, Scriven,
  and Davis}}]{Deitrick:92}
\bibinfo{author}{\bibfnamefont{G.~L.} \bibnamefont{Deitrick}},
  \bibinfo{author}{\bibfnamefont{L.~E.} \bibnamefont{Scriven}},
  \bibnamefont{and} \bibinfo{author}{\bibfnamefont{H.}~\bibnamefont{Davis}},
  \bibinfo{journal}{Molecular Simulation} \textbf{\bibinfo{volume}{8}},
  \bibinfo{pages}{239} (\bibinfo{year}{1992}).

\bibitem[{\citenamefont{Neumann}(1983)}]{Neumann:83}
\bibinfo{author}{\bibfnamefont{M.}~\bibnamefont{Neumann}},
  \bibinfo{journal}{Mol. Phys.} \textbf{\bibinfo{volume}{50}},
  \bibinfo{pages}{841} (\bibinfo{year}{1983}).

\bibitem[{\citenamefont{Roberts and Schnitker}(1994)}]{Roberts:94}
\bibinfo{author}{\bibfnamefont{J.~E.} \bibnamefont{Roberts}} \bibnamefont{and}
  \bibinfo{author}{\bibfnamefont{J.}~\bibnamefont{Schnitker}},
  \bibinfo{journal}{J. Chem. Phys.} \textbf{\bibinfo{volume}{101}},
  \bibinfo{pages}{5024} (\bibinfo{year}{1994}).

\bibitem[{\citenamefont{Roberts and Schnitker}(1995)}]{Roberts:95}
\bibinfo{author}{\bibfnamefont{J.~E.} \bibnamefont{Roberts}} \bibnamefont{and}
  \bibinfo{author}{\bibfnamefont{J.}~\bibnamefont{Schnitker}},
  \bibinfo{journal}{J. Phys. Chem.} \textbf{\bibinfo{volume}{99}},
  \bibinfo{pages}{1322} (\bibinfo{year}{1995}).

\bibitem[{\citenamefont{Durell and Wallqvist}(1996)}]{Durell:96}
\bibinfo{author}{\bibfnamefont{S.~R.} \bibnamefont{Durell}} \bibnamefont{and}
  \bibinfo{author}{\bibfnamefont{A.}~\bibnamefont{Wallqvist}},
  \bibinfo{journal}{Biophysical Journal} \textbf{\bibinfo{volume}{71}},
  \bibinfo{pages}{1695} (\bibinfo{year}{1996}).

\bibitem[{\citenamefont{Boresch and Steinhauser}(2001)}]{Boresch:2001}
\bibinfo{author}{\bibfnamefont{S.}~\bibnamefont{Boresch}} \bibnamefont{and}
  \bibinfo{author}{\bibfnamefont{O.}~\bibnamefont{Steinhauser}},
  \bibinfo{journal}{J. Chem. Phys.} \textbf{\bibinfo{volume}{115}},
  \bibinfo{pages}{10793} (\bibinfo{year}{2001}).

\bibitem[{\citenamefont{{Fernandez-Prini} and Crovetto}(1998)}]{Prini:89}
\bibinfo{author}{\bibfnamefont{R.}~\bibnamefont{{Fernandez-Prini}}}
  \bibnamefont{and} \bibinfo{author}{\bibfnamefont{R.}~\bibnamefont{Crovetto}},
  \bibinfo{journal}{J. Phys. Chem. Ref. Data} \textbf{\bibinfo{volume}{18}},
  \bibinfo{pages}{1231} (\bibinfo{year}{1998}).

\bibitem[{\citenamefont{Wagner and Pru{\ss}}(2002)}]{Wagner:2002}
\bibinfo{author}{\bibfnamefont{W.}~\bibnamefont{Wagner}} \bibnamefont{and}
  \bibinfo{author}{\bibfnamefont{A.}~\bibnamefont{Pru{\ss}}},
  \bibinfo{journal}{J. Phys. Chem. Ref. Data} \textbf{\bibinfo{volume}{31}},
  \bibinfo{pages}{387} (\bibinfo{year}{2002}).

\bibitem[{\citenamefont{Luzar and Chandler}(1996)}]{Luzar:96:1}
\bibinfo{author}{\bibfnamefont{A.}~\bibnamefont{Luzar}} \bibnamefont{and}
  \bibinfo{author}{\bibfnamefont{D.}~\bibnamefont{Chandler}},
  \bibinfo{journal}{Nature} \textbf{\bibinfo{volume}{379}}, \bibinfo{pages}{55}
  (\bibinfo{year}{1996}).

\bibitem[{\citenamefont{Geiger et~al.}(1986)\citenamefont{Geiger, Mausbach, and
  Schnitker}}]{Geiger:86}
\bibinfo{author}{\bibfnamefont{A.}~\bibnamefont{Geiger}},
  \bibinfo{author}{\bibfnamefont{P.}~\bibnamefont{Mausbach}}, \bibnamefont{and}
  \bibinfo{author}{\bibfnamefont{J.}~\bibnamefont{Schnitker}}, in
  \emph{\bibinfo{booktitle}{Water and Aqueous Solutions}}, edited by
  \bibinfo{editor}{\bibfnamefont{G.~W.} \bibnamefont{Neilson}}
  \bibnamefont{and} \bibinfo{editor}{\bibfnamefont{J.~E.}
  \bibnamefont{Enderby}} (\bibinfo{publisher}{Adam Hilger},
  \bibinfo{address}{Bristol}, \bibinfo{year}{1986}), pp.
  \bibinfo{pages}{15--30}.

\bibitem[{\citenamefont{Sciortino et~al.}(1991)\citenamefont{Sciortino, Geiger,
  and Stanley}}]{Sciortino:91}
\bibinfo{author}{\bibfnamefont{F.}~\bibnamefont{Sciortino}},
  \bibinfo{author}{\bibfnamefont{A.}~\bibnamefont{Geiger}}, \bibnamefont{and}
  \bibinfo{author}{\bibfnamefont{H.~E.} \bibnamefont{Stanley}},
  \bibinfo{journal}{Nature} \textbf{\bibinfo{volume}{354}},
  \bibinfo{pages}{218} (\bibinfo{year}{1991}).

\bibitem[{\citenamefont{Sciortino et~al.}(1992)\citenamefont{Sciortino, Geiger,
  and Stanley}}]{Sciortino:92}
\bibinfo{author}{\bibfnamefont{F.}~\bibnamefont{Sciortino}},
  \bibinfo{author}{\bibfnamefont{A.}~\bibnamefont{Geiger}}, \bibnamefont{and}
  \bibinfo{author}{\bibfnamefont{H.~E.} \bibnamefont{Stanley}},
  \bibinfo{journal}{J. Chem. Phys.} \textbf{\bibinfo{volume}{96}},
  \bibinfo{pages}{3857} (\bibinfo{year}{1992}).

\bibitem[{\citenamefont{Poole et~al.}(1992)\citenamefont{Poole, Sciortino,
  Essmann, and Stanley}}]{Poole:92}
\bibinfo{author}{\bibfnamefont{P.~H.} \bibnamefont{Poole}},
  \bibinfo{author}{\bibfnamefont{F.}~\bibnamefont{Sciortino}},
  \bibinfo{author}{\bibfnamefont{U.}~\bibnamefont{Essmann}}, \bibnamefont{and}
  \bibinfo{author}{\bibfnamefont{H.~E.} \bibnamefont{Stanley}},
  \bibinfo{journal}{Nature} \textbf{\bibinfo{volume}{360}},
  \bibinfo{pages}{324} (\bibinfo{year}{1992}).

\bibitem[{\citenamefont{Poole et~al.}(1993)\citenamefont{Poole, Sciortino,
  Essmann, and Stanley}}]{Poole:93:1}
\bibinfo{author}{\bibfnamefont{P.~H.} \bibnamefont{Poole}},
  \bibinfo{author}{\bibfnamefont{F.}~\bibnamefont{Sciortino}},
  \bibinfo{author}{\bibfnamefont{U.}~\bibnamefont{Essmann}}, \bibnamefont{and}
  \bibinfo{author}{\bibfnamefont{H.~E.} \bibnamefont{Stanley}},
  \bibinfo{journal}{Phys. Rev. E} \textbf{\bibinfo{volume}{48}},
  \bibinfo{pages}{3799} (\bibinfo{year}{1993}).

\bibitem[{\citenamefont{Ashbaugh et~al.}(2002)\citenamefont{Ashbaugh, Truskett,
  and Debenedetti}}]{Ashbaugh:2002}
\bibinfo{author}{\bibfnamefont{H.~S.} \bibnamefont{Ashbaugh}},
  \bibinfo{author}{\bibfnamefont{T.~M.} \bibnamefont{Truskett}},
  \bibnamefont{and} \bibinfo{author}{\bibfnamefont{P.~G.}
  \bibnamefont{Debenedetti}}, \bibinfo{journal}{J. Chem. Phys.}
  \textbf{\bibinfo{volume}{116}}, \bibinfo{pages}{2907} (\bibinfo{year}{2002}).

\bibitem[{\citenamefont{Starr et~al.}(1999)\citenamefont{Starr, Nielsen, and
  Stanley}}]{Starr:99}
\bibinfo{author}{\bibfnamefont{F.~W.} \bibnamefont{Starr}},
  \bibinfo{author}{\bibfnamefont{J.~K.} \bibnamefont{Nielsen}},
  \bibnamefont{and} \bibinfo{author}{\bibfnamefont{H.~E.}
  \bibnamefont{Stanley}}, \bibinfo{journal}{Phys. Rev. Lett.}
  \textbf{\bibinfo{volume}{82}}, \bibinfo{pages}{2294} (\bibinfo{year}{1999}).

\end{thebibliography}

\end{document}